\documentclass[11pt]{article}
\pdfoutput=1
\usepackage[margin=1.0in]{geometry}
\usepackage{amsmath,amssymb,graphicx}
\usepackage{dsfont}
\usepackage{xcolor}
\definecolor{Mahogany}{rgb}{0.62,0.24,0.15}
\definecolor{DarkRed}{rgb}{0.6,0,0}
\definecolor{DarkGreen}{rgb}{0,0.6,0}
\definecolor{DarkBlue}{rgb}{0,0,0.6}
\definecolor{gray}{RGB}{128,128,128}
\usepackage[colorlinks=true,linktocpage=true,linkcolor=DarkBlue,citecolor=DarkGreen,urlcolor=DarkGreen]{hyperref}
\usepackage{slashed}
\usepackage{verbatim}
\usepackage{cite}

\usepackage{setspace}
\usepackage[T1]{fontenc}

\usepackage[utf8]{inputenc}

\usepackage{upgreek}
\usepackage{bbm} 

\usepackage[font=small,labelfont=bf]{caption}
\let\OLDthebibliography\thebibliography
\renewcommand\thebibliography[1]{
    \small
    \OLDthebibliography{#1}
    \setlength{\parskip}{0.0pt plus 0.5pt}
    \setlength{\itemsep}{3.5pt plus 2.0pt minus 1.0pt}
}

\usepackage{multirow}

\newcommand{\lab}[1]{{\mathrm{#1}}}

\def\e{{\lab{e}}}

\DeclareMathOperator{\Tr}{Tr}
\DeclareMathOperator{\diag}{diag}

\renewcommand{\Im}{\mathrm{Im}}

\def\dd{\text{d}}
\newcommand{\hc}{{\textrm{h.c.}}}

\def\bsigma{\bar{\sigma}}

\newcommand{\minus}{{\scalebox {0.75}[1.0]{$-$}}}

\newcommand{\be}{\begin{equation}}
\newcommand{\ee}{\end{equation}}
\newcommand{\bp}{\begin{pmatrix}}
\newcommand{\ep}{\end{pmatrix}}
\newcommand{\ba}{\begin{aligned}}
\newcommand{\ea}{\end{aligned}}

\def\phisd{\phi_{\textsc{sd}}}

\def\phidt{\phi_{\textsc{dt}}}

\title{\bf Barr--Zee Diagrams at a High-Energy Muon Collider}
\author{
Samuel Homiller$^1$,
Jackie Lodman$^2$,
Aditya Parikh$^3$ and
Matthew Reece$^2$\\[0.5em]
{\small \color{gray} \texttt{shomiller@cornell.edu},~
\texttt{j.lodman@g.harvard.edu},}\\[0.em]
{\small \color{gray} \texttt{aditya.parikh@stonybrook.edu},~
\texttt{mreece@g.harvard.edu}}\\[0.5em]
{\small${}^1$Laboratory for Elementary Particle Physics, Cornell University, Ithaca, NY 14853}\\[0.25em]
{\small${}^2$Department of Physics, Harvard University, Cambridge, MA 02138}\\[0.25em]
{\small${}^3$C.\,N. Yang Institute for Theoretical Physics, Stony Brook University, Stony Brook, NY 11794}
}

\begin{document}

\maketitle

\vspace*{-2em}

\begin{abstract}
The sensitivity of electron EDM experiments has been increasing at a rapid pace, and could yield indications of new physics in the coming decade. An intriguing possibility is that an EDM signal could be generated by new, electroweak-charged particles at the TeV scale that couple to the Higgs and contribute to the electron EDM at two-loop order via Barr--Zee diagrams. A high-energy muon collider could decisively search for new physics at this scale. In this work, we explore this complementarity between colliders and EDM experiments, and note that Barr--Zee diagrams from the aforementioned particles are closely related to vector-boson scattering processes at a muon collider. These loop corrections lead to kinematic features in the differential cross sections of these processes, dictated by the optical theorem. We demonstrate this connection in the context of the singlet-doublet and doublet-triplet extensions to the SM, explore the detectability of these features at a muon collider experiment, and discuss how these measurements can be used to ascertain the underlying model parameters.
\end{abstract}

\vskip 0.5cm

\tableofcontents
\vskip 0.5cm

\section{Introduction}

While the Standard Model (SM) has thus far proven remarkably resilient to experimental scrutiny, the existing theoretical motivations for new physics, as well as forthcoming advances in experimental sensitivity, provide ample reason to be optimistic that this situation is only temporary.
In particular, the next decade promises significant improvements in a number of observables, such as searches for lepton-flavor violation and CP-violating electric dipole moments (EDMs) of SM particles~\cite{Davidson:2022jai, Alarcon:2022ero}. 
These observables are special because they are not explicitly forbidden by the SM symmetries, but they are extremely suppressed by the particular flavor structure of the Standard Model. 
This makes them excellent probes for new physics, as a measurement of a nonzero value at the near-future experimental sensitivity would be a clear smoking gun signal of BSM physics. 

Of these probes, the electron EDM is especially exciting. The current 90\% C.L. bound $|d_e| < 4.1 \times 10^{-30}\,e\cdot \textrm{cm}$ set by the JILA experiment \cite{Roussy:2022cmp} naively sets indirect constraints on new sources of CP violation at scales as high as $75~\textrm{TeV}$, depending on how the CP-violation is communicated to the SM~\cite{Cesarotti:2018huy}. 
While the sensitivity of these experiments is remarkable, they are still several orders of magnitude away from probing the known sources of CP-violation in the SM.
The phases in the lepton sector lead to EDMs of order $|d_e| \lesssim 10^{-43}\,e\cdot \textrm{cm}$.
The strong CP phase $\bar\theta$ leads to an EDM $|d_e| \lesssim 10^{-37}\,e\cdot \textrm{cm}$ \cite{Choi:1990cn, Ghosh:2017uqq}.
The CKM phase generates an EDM at the four-loop level, estimated to be $|d_e| \sim 10^{-44}\,e\cdot \textrm{cm}$~\cite{Pospelov:2013sca, Hoogeveen:1990cb, Pospelov:1991zt}. 
Both the CKM phase and the strong CP phase also generate a CP-odd electron-nucleon coupling, which can mimic an EDM signal in experiments as large as $|d_e| \sim 10^{-35}\,e\cdot \textrm{cm}$~\cite{Ema:2022yra} or $10^{-32}\,e\cdot\textrm{cm}$~\cite{Flambaum:2019ejc}, respectively. Disentangling the CP-odd electron-neutron couplings from a genuine electron EDM requires performing measurements in multiple atomic or molecular species. 
Given all this, it is not overly optimistic to suppose that we may receive the first indirect signal of physics beyond the SM in the near future.

Even though a nonzero $|d_e|$ measurement above the SM expectations would be a smoking gun signal for new physics, it would tell us very little about the underlying nature of the new physics and the associated mass scale. One possibility is that the new CP-violating physics couples directly to the electron, leading to one-loop EDMs, which is common in supersymmetric extensions of the SM. Another, more generic, scenario is that the EDM is induced by new particles carrying electroweak quantum numbers and interacting with the Higgs. In fact, any such particle will induce an electron EDM at the two-loop level through what are now known as ``Barr-Zee diagrams'' \cite{Barr:1990vd}. This is particularly exciting because the current sensitivity places the mass scale of two-loop new physics contributions to the electron EDM at $\mathcal{O}(\textrm{few TeV})$. 
This possibility makes it crucial that the next-generation collider experiment be able to decisively test new electroweak physics at the TeV scale. In this light, a high-energy muon collider provides an enticing opportunity.

The possibility of a muon collider operating at multi-TeV center of mass energies has received a great deal of attention in the past few years~\cite{AlAli:2021let, Aime:2022flm, MuonCollider:2022nsa, Black:2022cth}. 
While there are numerous challenges in producing intense, high-energy muon beams suitable for collisions, no insurmountable obstacles have been identified, and there is now a concerted effort to develop the necessary accelerator technology and develop a realistic machine design~\cite{MuonCollider:2022glg, MuonCollider:2022ded, Bartosik:2024ulr, Skoufaris:2024okx, InternationalMuonCollider:2024jyv}. 
This effort is galvanized by the excitement around the physics potential such a machine would offer. 
Among the numerous physics opportunities featured in the reports above, a muon collider would have impressive capabilities in precision Higgs/electroweak physics~\cite{Chiesa:2020awd, Han:2020pif, Buttazzo:2020uzc, Forslund:2022xjq, Forslund:2023reu, Han:2023njx, Li:2024joa}, searches for new heavy states~\cite{Buttazzo:2018qqp, Asadi:2021gah, Bao:2022onq, Han:2022mzp, Li:2023tbx, Asadi:2023csb}, and searches that are complementary to forthcoming precision probes~\cite{Homiller:2022iax, Asadi:2024lfv}. 
This is in addition to the exciting possibilities with parasitic facilities alongside a collider/accelerator complex~\cite{Cesarotti:2022ttv, Cesarotti:2023sje, Bogacz:2022xsj, Gori:2024zbs}. 

The goal of this manuscript is to highlight a completely different type of signature. The new electroweak states responsible for an EDM would also lead to loop corrections to vector-boson scattering processes at a muon collider, via diagrams that are closely related to the Barr--Zee diagrams mentioned above. As a result of the optical theorem, these loop corrections lead to a kinematic feature in the partonic center-of-mass energy, which may be identified in a muon collider experiment. 
This signature---likely accessible only at a high-energy lepton machine---would provide conclusive evidence of the relationship between an observed electron EDM and new electroweak states discoverable at a muon collider.
Moreover, as we discuss, it may offer crucial information into the underlying parameters of the high-energy theory, orthogonal to direct searches and EDM measurements. 

Similar features that result from unitarity cuts have been proposed as a potential signal in other contexts, e.g., as a dark matter signal in~\cite{Altmannshofer:2014cla}, and as a possible effect in the diphoton invariant mass spectrum in~\cite{Chway:2015lzg}. Here, the process of interest, underlying model and interpretation are all different.

\medskip

The rest of this paper is structured as follows: 
In Section~\ref{sec:models}, we introduce two simplified models with new electroweak states that lead to electron EDMs via two-loop Barr--Zee diagrams. In Section~\ref{sec:collider_constraints} we review the current constraints and future prospects for discovering new electroweak states directly or indirectly at colliders, and discuss how these measurements inform us about the parameters of the underlying theories. Our main results are presented in Section~\ref{sec:vbf}, where we demonstrate the presence of the kinematic features in vector-boson scattering from one-loop corrections due to new electroweak states, and demonstrate how these could be discovered at a muon collider and used as a measurement of the model parameters. We conclude in Section~\ref{sec:conclusion}, and present a few additional details of our statistical analysis in Appendix~\ref{app:likelihood_details}.

\section{Minimal Models of Higgs-Coupled CP Violation}
\label{sec:models}

As discussed in the Introduction, a nonzero electron EDM would indicate a new source of CP violation, and hence, new physics.
While new CP-violating phases are ubiquitous in physics beyond the SM, a particularly interesting possibility is that this new source of CP-violation is associated with the physics of electroweak symmetry breaking near the TeV scale and the Higgs boson.
This scenario arises in various extensions to the Standard Model, most notably supersymmetry. 
For our purposes, however, it will be useful to identify models with the minimal field content necessary to construct this scenario, and study their signatures.
Our conclusions from these simplified models can be straightforwardly extrapolated to more specific scenarios, and illustrate the general point.

To this end, we seek models with new electroweak multiplets, which must be in representations whose product contains an $\lab{SU}(2)_L$ doublet to permit coupling to the Higgs. For simplicity, we focus on color-neutral fermions and examples with the minimal field content allowing mass terms for the new fermions.
These considerations lead us to two minimal, benchmark models: the singlet-doublet and doublet-triplet fermion models. As we will review below, each of these models introduces a single additional CP-violating phase, with slightly different phenomenology associated with the Higgs boson.

\subsection{The Singlet-Doublet Model}

The singlet-doublet model~\cite{Mahbubani:2005pt, DEramo:2007anh, Enberg:2007rp, Cohen:2011ec, Cheung:2013dua, Abe:2014gua, Calibbi:2015nha, Freitas:2015hsa, Banerjee:2016hsk, Cai:2016sjz, LopezHonorez:2017zrd, Fraser:2020dpy} extends the SM with an $\lab{SU}(2)_L$ singlet fermion $\xi$  with vanishing hypercharge, along with a pair of $\lab{SU}(2)_L$ doublets $\psi_u$ and $\psi_d$ with hypercharges $\pm 1/2$, respectively.
We'll write the $\lab{SU}(2)_L$ components of $\psi_u$ and $\psi_d$ as
\begin{equation}
\psi_u = \bp \psi^+ \\ \psi_u^0 \ep \, , \qquad
\psi_d = \bp \psi_d^0 \\ \psi^- \ep \, .
\end{equation}
This model has been studied extensively in the context of Higgs-portal dark matter, and also arises as the bino-higgsino sector of the MSSM in the limit where the wino mass $M_2 \to \infty$. 

These charge assignments allow for a Majorana mass for $\xi$ and a Dirac mass for $\psi_u$ and $\psi_d$.
They also allow for Yukawa interactions with the Higgs, so that the full Lagrangian takes the form:\footnote{
We are writing all the fermion fields as left-handed Weyl spinors, using the conventions of~\cite{Dreiner:2008tw}. When necessary, we'll denote $\lab{SU}(2)_L$ indices as $a, b, \dots$. When not written explicitly, they are assumed to be contracted in the ``natural'' way, with two fundamentals (anti-fundamentals) contracted via $\epsilon^{ab}$ ($\epsilon_{ab}$), with $\epsilon^{12} = \epsilon_{21} = +1$, and a fundamental and anti-fundamental contracted with the Kronecker delta-symbol, $\delta^a_b$.}
\begin{equation}
\label{eq:uv_lagrangian_sd}
\begin{aligned}
\mathcal{L} & = \mathcal{L}_{\textrm{SM}} 
+ i \xi^{\dagger} \bsigma^{\mu} \partial_{\mu} \xi
+ i \psi_u^{\dagger} \bsigma^{\mu} D_{\mu} \psi_u
+ i \psi_d^{\dagger} \bsigma^{\mu} D_{\mu} \psi_d \\[0.25em]
& \qquad 
- \Big(\,
\frac{1}{2} M_1 \xi \xi 
+ \mu \psi_u \psi_d 
- \tilde{Y}_u \xi H^{\dagger} \psi_u
+ \tilde{Y}_d \xi H \psi_d 
+ \hc \Big)
\end{aligned}
\end{equation}
This Lagrangian contains 3 new fields and 4 parameters which are generically complex. Field rotations leave us with one physical CP-violating phase, which we choose to distribute in the Yukawas such that
\begin{equation}
\label{eq:Y_phases}
\tilde{Y}_u = Y_u\, \e^{i\frac{\phisd}{2}}, \qquad 
\tilde{Y}_d = Y_d\, \e^{i\frac{\phisd}{2}}
\end{equation}
where $Y_u, Y_d \in \mathbb{R}$. There is also a $\mathbb{Z}_{2}$ symmetry where $Y_u \to \minus Y_u$, $Y_d \to \minus Y_d$, and $\xi \to \minus \xi$ that leaves the Lagrangian invariant, so we will restrict our attention to $Y_{u} \geq 0$ for the remainder of our analysis.
    
After electroweak symmetry breaking, this model consists of a single $\lab{U}(1)_{\textsc{em}}$-charged Dirac fermion, $\psi$, and three neutral Majorana fermions that we denote $\chi$. We refer to these as the charginos and neutralinos, respectively, in analogy to the MSSM.

To understand the mass spectrum, we package the three neutral fermions into $\tilde{\chi} = (\psi_u~~\psi_d~~\xi)^T$, so that
\begin{equation}
\mathcal{L} \supset
- \mu \psi^+ \psi^- 
- \frac{1}{2} \tilde{\chi}_i M^{ij} \tilde{\chi}_j
+ \hc + \dots
\end{equation}
where the mass matrix takes the form
\begin{equation}
M = \bp
0 & \minus \mu & \minus \frac{1}{\sqrt{2}} \tilde{Y}_u v \\
\minus \mu & 0 & \minus \frac{1}{\sqrt{2}} \tilde{Y}_d v \\
\minus \frac{1}{\sqrt{2}} \tilde{Y}_u v & \minus \frac{1}{\sqrt{2}} \tilde{Y}_d v & M_1 
\ep \, .
\end{equation}
To move to the mass basis, we diagonalize the mass-squared matrix $M^{\dagger}M$ via a unitary matrix $U$, with the eigenvectors of $M^{\dagger}M$ as its columns. This diagonalizes $M$, and we can simultaneously rotate the individual eigenvectors such that $M$ has real, positive eigenvalues. Explicitly, 
\begin{equation}
U^{\dagger} M^{\dagger}M U = \diag(m_{\chi_1}^2, m_{\chi_2}^2, m_{\chi_3}^2)\,, \qquad
U^T M U = \diag(m_{\chi_1}, m_{\chi_2}, m_{\chi_3})\,, 
\end{equation}
with $m_{\chi_1}$, $m_{\chi_2}$, $m_{\chi_3}$ chosen to be real and nonnegative. We label the corresponding mass eigenstate neutral fermions as $\chi = (\chi_1~~\chi_2~~\chi_3)^T$, defined by $\tilde{\chi}_j = U_j{}^k \chi_k$, with the index assumed to be mass-ordered.

Note that $\lab{SU}(2)_L$ invariance in this model does not allow for a renormalizable term coupling the neutral Higgs boson to the charged fermion: the only Higgs coupling is to the neutralinos. This is in stark contrast to the doublet-triplet model discussed below, and will have a significant impact on the phenomenology, as we will discuss. 

When $Y_u = \minus Y_d$, the singlet-doublet interactions are invariant under the approximate $\lab{SU}(2)$ custodial symmetry of the SM Higgs sector. In this limit, one of the neutralino-Higgs interactions vanishes, while the other is maximized. The mass spectrum can also be solved analytically---for details, see~\cite{Freitas:2015hsa}.

\subsection{The Doublet-Triplet Model}

We next consider the doublet-triplet model~\cite{Dedes:2014hga, Abe:2014gua, Freitas:2015hsa, LopezHonorez:2017zrd, Fraser:2020dpy}.
This model similarly arises in the MSSM, but instead as the higgsino-wino sector, taking the $M_1 \to \infty$ limit.

The model consists of the same $\lab{SU}(2)_L$ doublets $\psi_u$ and $\psi_d$ as above, now alongside an $\lab{SU}(2)_L$ triplet of $Y = 0$ fermions, $\lambda_A$. Here, $A = 1 \dots 3$ is an adjoint $\lab{SU}(2)_L$ index. It's convenient to expand the components of the triplet as
\begin{equation}
(\lambda)_a{}^b \equiv \frac{1}{\sqrt{2}} \lambda_A (\sigma^A)_a{}^b 
= \bp \frac{1}{\sqrt{2}} \lambda^0 & \lambda^+ \\ 
\lambda^- & \minus \frac{1}{\sqrt{2}}\lambda^0 \ep \,.
\end{equation}

The Lagrangian for the doublet-triplet model takes the form
\begin{equation}
\label{eq:uv_lagrangian_dt}
\begin{aligned}
\mathcal{L} & = \mathcal{L}_{\textrm{SM}} 
+ i \lambda_A^{\dagger} \bsigma^{\mu} D_{\mu} \lambda^A
+ i \psi_u^{\dagger} \bsigma^{\mu} D_{\mu} \psi_u
+ i \psi_d^{\dagger} \bsigma^{\mu} D_{\mu} \psi_d \\[0.25em]
& \qquad 
- \Big(\,
\frac{1}{2} M_2 \Tr \lambda\,\lambda
+ \mu\, \psi_u \psi_d 
+ \tilde{Y}_1 H^{\dagger} \lambda \psi_u 
+ \tilde{Y}_2 \epsilon^{ab} H_a (\lambda)_b{}^c \psi_{d\,c}
+ \hc \Big)\,,
\end{aligned}
\end{equation}
where we have written the $\lab{SU}(2)_L$ indices in the last term explicitly for clarity. As before, this model also contains 3 new fields and 4 new complex parameters, so we are left with one physical CP-violating phase which we fix to be in the Yukawas, in complete analogy to Eq.~\eqref{eq:Y_phases}.
This model also contains a custodial symmetric limit when $Y_1 = \minus Y_2$, as discussed in~\cite{Dedes:2014hga, Freitas:2015hsa}.

In the IR, this model consists of three neutral Majorana fermions, as in the singlet-doublet model, but now {\em two} charged Dirac fermions. We write,
\begin{equation}
\tilde{\chi} = \bp \psi_u^0 \\[0.25em] \psi_d^0 \\[0.25em] \lambda^0 \ep \,, \qquad 
\Psi^+ = \bp \psi^+ \\ \lambda^+ \ep \,, \qquad
\Psi^- = \bp \psi^- \\ \lambda^- \ep \,.
\end{equation}
In terms of these, the mass terms in the Lagrangian can be written,
\begin{equation}
\mathcal{L} \supset 
-\Psi_i^+ (M_c)^i{}_j \Psi^{-\,j}
-\frac{1}{2} \tilde{\chi}_i (M_n)^{ij} \tilde{\chi}_j
+ \hc + \dots
\end{equation}
where
\begin{equation}
M_c = \bp
\mu & \frac{1}{\sqrt{2}} \tilde{Y}_1 v \\
\minus \frac{1}{\sqrt{2}} \tilde{Y}_2 v & M_2 \ep \,, \qquad
M_n = \bp
0 & \minus \mu & \minus\frac{1}{2} \tilde{Y}_1 v \\
\minus \mu & 0 & \minus\frac{1}{2} \tilde{Y}_2 v \\
\minus \frac{1}{2} \tilde{Y}_1 v & \minus \frac{1}{2} \tilde{Y}_2 v & M_2 
\ep \, .
\end{equation}
The diagonalization of the neutral mixing matrix is completely analogous to the singlet-doublet model. We denote the corresponding unitary matrix as $U_n$, with the mass eigenstates defined as $\tilde{\chi}_j = (U_n)_j{}^k \chi_k$ having masses $m_{n,k}$. 

For the charged fermions, $\Psi^+$ and $\Psi^-$ each represent two independent fermion degrees of freedom. The mass matrix is diagonalized by two independent rotations for the positively and negatively charged fields. We introduce matrices $U_c$ and $V_c$ satisfying
\begin{equation}
\label{eq:umix_cha_dt}
\begin{gathered}
V_c^T M_c U_c = \diag(m_{\psi_1}, m_{\psi_2}), \\[0.25em]
U_c^{\dagger} M_c^{\dagger}M_c U_c = V_c^T M_c M_c^{\dagger} V_c^* = \diag(m_{\psi_1}^2, m_{\psi_2}^2)\, ,
\end{gathered}
\end{equation}
where the $m_{\psi_i}$ are real and nonnegative. The mass eigenstate spinors are denoted $\psi^+_i$ and $\psi^-_j$, $i,j = 1,2$, and defined by
\begin{equation}
\Psi^+_i = (V_c)_i{}^k \psi^+_k \, \qquad
\Psi^{-\,j} = (U_c)^j{}_l \psi^{-\,l}\, .
\end{equation}
In contrast to the singlet-doublet model, the charginos couple to the Higgs in this setup which leads to more interesting phenomenology. 

\subsection{Contributions to the Electron EDM}

New sources of CP-violation will manifest as contributions to the electron EDM. The current experimental upper bound on the EDM is placed at $|d_e| < 4.1 \times 10^{-30}\,e\cdot \textrm{cm}$ from JILA~\cite{Roussy:2022cmp}. The next generation of experiments will tighten these constraints even further with the hopes of detecting an electron EDM above the expected Standard Model value of $|d_e| \sim 10^{-35}\,e\cdot \textrm{cm}$ as predicted in~\cite{Ema:2022yra}.
Any discovery of a nonzero EDM above this scale points towards new physics, which could easily reside at the multi-TeV scale, assuming the contribution comes via two-loop Barr--Zee-type diagrams.\footnote{As noted earlier, the field content of the singlet-doublet and doublet-triplet models is similar to the Bino-Higgsino or Wino-Higgsino sectors of the MSSM, assuming the other superpartners decouple. This is the case for instance in ``mini-split'' supersymmetry~\cite{Giudice:1998xp, Wells:2003tf, Wells:2004di, Giudice:2011cg, Arvanitaki:2012ps, Arkani-Hamed:2012fhg}, in which case the one-loop contributions to the EDM from squarks and sleptons are suppressed relative to the two-loop electroweak contributions discussed below.}
If the CP-violating couplings to the electron are more direct, the scale of new physics can be even higher~\cite{Cesarotti:2018huy,Panico:2018hal}.

\begin{figure}[t]
\centering
\includegraphics[scale=1.2]{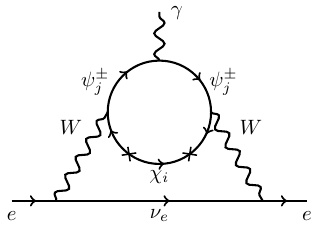}\qquad
\includegraphics[scale=1.2]{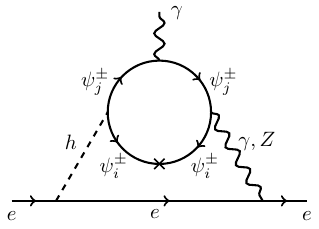}
\caption{These are the two-loop Barr--Zee diagrams involving new fermions that contribute to the electron EDM. $\psi_j^\pm$ denote the charginos and $\chi_i$ denote the neutralinos. In the singlet-doublet model, the charginos don't couple to the Higgs and, as a result, the dominant contribution to the EDM comes from the diagram on the left. In the doublet-triplet model, since the charginos also couple to the Higgs, we get contributions from both diagrams. In the diagram on the right, $i \neq j$ only contributes for the $Z$ boson leg, while $i = j$ contributes in both cases.
}
\label{fig:edm_diagrams}
\end{figure}

The presence of BSM states carrying electroweak charges, and the absence of direct couplings with the electrons, means that the leading contribution to the electron EDM comes in the form of two-loop Barr--Zee diagrams~\cite{Barr:1990vd}. For the singlet-doublet model, the only relevant diagram comes with $W$ boson legs, as shown in the left panel of Fig.~\ref{fig:edm_diagrams}. Taking the neutrinos as massless, approximating the lepton couplings as flavor diagonal, and recognizing that one of the internal loop fermions is neutral, we can compute the Barr--Zee diagram using a simplified version of Equation 21 in~\cite{Atwood:1990cm} (see also~\cite{Kadoyoshi:1996bc, Chang:2005ac}),
\begin{equation}
    \begin{split}
    \frac{d_e}{e} = -\frac{g^2}{(4 \pi)^4}\sum_{i} \text{Im}([aU]_{i}^* [bU^*]_{i})\left(\frac{\mu m_{n,i} m_e}{M_W^4}\right)G\left(\frac{\mu^2}{M_W^2}, \frac{m_{n,i}}{M_W^2}, 0\right),
    \end{split}
\end{equation}
where $a$ and $b$ parameterize the left- and right-handed $W$-boson couplings to the inner loop fermions. $G(\alpha, \beta, \gamma)$ is defined as
\begin{equation}
   G(\alpha, \beta, \gamma) = \frac{1}{1-\gamma}\int_0^1 \frac{dx}{1 - x} \left( \frac{\gamma}{z - \gamma} \log\left(\frac{\gamma}{z}\right) + \frac{1}{1 - z} \log \left( \frac{1}{z}\right)\right)
\end{equation}
with
\begin{equation}
    z(x,\alpha,\beta) = \frac{\beta}{x} + \frac{\alpha}{1 - x}.
\end{equation}

The doublet-triplet model contains two additional contributions shown in the right panel of Figure~\ref{fig:edm_diagrams}.\footnote{Additional diagrams with this topology also exist, but we can argue that their contributions to the electron EDM are negligibly small or vanishing as discussed in~\cite{Fraser:2020dpy}. The $\gamma\gamma$ diagram does not contribute due to Furry's theorem which states that pure QED amplitudes with an odd number of external photons vanish. The $ZZ$ diagram does not contribute since the $Z$-boson is a real boson which does not have a 1-loop EDM as argued in~\cite{Atwood:1990cm}. The $\gamma Z$ diagram contribution vanishes since the same charged fermion runs through the entire loop and CP-violation can't enter since the diagonal $Z$-couplings are real. Finally, the $hh$ diagram contributes but is suppressed by additional factors of the electron Yukawa.}
The general form of these contributions from~\cite{Nakai:2016atk} (see also earlier work on electroweakino EDM contributions, e.g.,~\cite{Pilaftsis:2002fe, Giudice:2005rz}) is given by
\begin{equation}
    d_{e}^{hV} = \frac{1}{16\pi^{2}m_{h}^{2}}\int_{0}^{1}dx\frac{1}{x(1-x)}j\Bigg(\frac{m_{V}^{2}}{m_{h}^{2}}, \frac{\tilde{\Delta}^{V}}{m_{h}^{2}}\Bigg)g_{e}^{V}c_{O}^{V}\frac{m_e}{v},
\end{equation}
where $v$ is the Higgs vev, the electron coupling to the $Z$ or $\gamma$ is denoted $g_e^V$, and $j(r,s)$ is given by
\begin{equation}
    j(r,s) = \frac{1}{r-s}\Bigg(\frac{r \log r}{r-1} - \frac{s\log s}{s-1}\Bigg).
\end{equation}

The inner loops, containing only charged fermions for $\gamma h$ and $Z h$, determine $c_{O}^{V}$ and $\tilde{\Delta}^{V}$, which are given by
\begin{equation}
\begin{split}
    c_O^{Z} = -\frac{e}{2 \pi^2} \text{Re}\Big( m_c^i x^2 (1 - x)(g_{ij}^Sg_{ji}^{V*} + i g_{ij}^P g_{ji}^{A*}) + (1 - x)^3 m_c^j (g_{ij}^S g_{ji}^{V*} - i g_{ij}^P g_{ji}^{A*})\Big), \\
    \tilde{\Delta}^{Z} = \frac{x m_c^i + (1 - x) m_c^j}{x (1 - x)}, \qquad \quad c_O^{\gamma} = - \frac{e^2 g^P_{jj}}{2 \pi ^2}(1 - x) m_c^j, \qquad \quad \tilde{\Delta}^{\gamma} = \frac{(m_c^{j})^2}{x (1 - x)},
\end{split}
\end{equation}
and the couplings are given by
\begin{equation}
    \begin{split}
        g^S &= \frac{1}{2}( U_c^T Y_c V_c + V_c^\dag Y_c^\dag U_c^*), \qquad g^P = \frac{i}{2}(U_c^T Y_c V_c - V_c^\dag Y_c^\dag U_c^*), \\
        g^V &= U_c^T U_+ U_c^* + V_c^\dag U_+ V_c,  \qquad \quad g^A =  U_c^T U_+ U_c^* - V_c^\dag U_+ V_c.
    \end{split}
\end{equation}
$Y_c$ parameterizes the coupling of the Higgs and $U_+$ encodes the couplings of the $Z$ to the internal charged fermions.
We also choose $\psi_j^\pm$ to be the fermion which radiates the on-shell external photon, and $g_{ij}^* = (g_{ji})^*$.

\section{Collider Constraints: Present and Future}
\label{sec:collider_constraints}

Beyond the constraints from the electron EDM discussed above, there are both direct and indirect constraints on new electroweak fermions from collider experiments. In this section, we will review these constraints, and discuss the projected discovery reach of a future high-energy muon collider. 
If the lightest neutral fermion is the dark matter, there are also constraints from direct detection experiments. We do not consider these as they rely on additional assumptions about the stability of the neutralinos and the calculation of the relic abundance (see, e.g.,~\cite{Arcadi:2017kky,Roszkowski:2017nbc} for additional details).

\subsection{Direct Searches at Colliders}
\label{subsec:direct_searches}

LEP and the LHC have both undertaken extensive direct searches for new electrically charged fermions. For charged fermions with masses $\lesssim 100~\textrm{GeV}$, LEP places the most stringent constraints, with the precise bound depending on the assumptions about the model~\cite{Egana-Ugrinovic:2018roi}. Assuming a small mass splitting between the charged and neutral states, LEP rules out charginos below 92.4 GeV. For mass splittings $\gtrsim$ 3 GeV, the constraints are stronger and extend to charginos up to $103.5~\textrm{GeV}$~\cite{LEPSUSYWG1,LEPSUSYWG2}. 

At the LHC---and at future high-energy colliders---the relevant searches and the resulting constraints depend significantly on the mass spectrum.
If the mass splitting between the lightest charged and neutral state is $\gtrsim 100~\textrm{GeV}$, $\psi^{\pm}$ decays to $\chi_1$ and an on-shell $W^{\pm}$, the latter of which has decays that can be readily detected.
In the opposite limit, when $m_{\psi^{\pm}} = m_{\chi_1}$, the chargino may be long-lived, with signatures such as charged tracks. For intermediate mass splittings, chargino decays through a virtual $W$ boson may still be prompt, but the energy in the decay products is small enough that they are nearly impossible to detect.

Large splittings, with readily detectable $W$ decay products, are generic in the singlet-doublet model when $M_1 \lesssim \mu$. 
For $M_1 \ll \mu$, the lightest neutralino is predominantly the $\lab{SU}(2)_L$ singlet, with mass governed by $M_1$ and $Y_{u,d} v$, while the chargino has mass $\mu$. 
For $M_1 \gtrsim \mu$, when $\chi_{1,2}$ are predominantly higgsino-like, the splitting effects depend on the size of the Yukawas. Working in the large $M_1$ limit, the splitting comes from a dimension-5 operator, and goes like
\begin{equation}
m_{\psi^{\pm}} - m_{\chi_1} \sim \frac{(Y_u - Y_d)^2 v^2}{M_1}\,.
\end{equation}
Depending on the relative sizes of $Y_{u,d} v$ and $\mu$, this splitting may be large enough so that the $W$ decay products are observable. If $Y_{u,d}$ are small (or equal), the splitting vanishes at tree-level. 
In our later analyses, we will focus our attention on the case $M_1 < \mu$, for simplicity.

In the doublet-triplet model, we can consider similar limiting cases.
When $M_2 \gg \mu$, the situation is completely analogous to the light higgsino-like case in the singlet-doublet model. 
In the $\mu \gg M_2$ limit, on the other hand, the lightest neutral state is wino-like. In this case, the light triplet still comes with a nearly-degenerate chargino, but the leading operator that can split the charged and neutral states is dimension-7. The leading correction to the splitting arises from loop corrections, and is $\mathcal{O}(100~\textrm{MeV})$~\cite{Ibe:2012sx}, leading to long-lived chargino signatures.
Interpolating between these limiting cases, we see that the doublet-triplet model always leads to a relatively compressed spectrum between the lightest charged and neutral states.\footnote{These scaling arguments break down if all the mass parameters in the doublet-triplet model are comparable to $v$. In this case, more exotic spectra are possible~\cite{Dedes:2014hga}. For the TeV-scale masses of interest to us, however, such cases are not possible.}

There are numerous searches at the LHC that place constraints beyond the LEP bound mentioned above. 
For a recent review compiling the different searches by ATLAS and CMS can be found in \cite{Canepa:2020ntc}. An even more recent result by CMS sets constraints on higgsino and wino production~\cite{CMS:2024gyw}. These searches can extend up to $m_{\psi} \approx 800~\textrm{GeV}$ ($1~\textrm{TeV}$) for the higgsino-like (wino-like) case when the lightest (bino-like) neutralino is very light, but are significantly weaker for more general neutralino masses.
Other relevant searches, with various assumptions about the mass spectrum, can be found in \cite{ATLAS:2021yqv, ATLAS:2022hbt, ATLAS:2022zwa}. 
These bounds will improve with the increased luminosity at the HL-LHC, but the qualitative picture is likely to remain the same: there is ample allowed parameter space for new electroweak states in the few hundred GeV to TeV mass range.

\medskip 

We now turn to the prospects of these searches at a high-energy muon collider. We will confine our attention to a 10 TeV center-of-mass energy collider, assuming a total integrated luminosity of $10~\textrm{ab}^{-1}$.
This approximates the ``site-filling'' design proposed at Fermilab~\cite{Black:2022cth}, and is frequently used as a benchmark in physics studies, providing a useful comparison. Most of our results would apply equally for any high-energy lepton collider; we focus on muons because the prospects for reaching the 10 TeV scale in the next generation of collider experiments appear more viable.

The capabilities of a $10\,\textrm{TeV}$ muon collider for discovering new electroweak-charged states are immense. 
As discussed in~\cite{AlAli:2021let}, provided the final states are distinctive enough, the charged particle-antiparticle pair-production cross section is large enough to facilitate discovery for masses $m_{\psi^{\pm}} \lesssim 0.9 \times (\sqrt{s}/2)$. This is clearly the case when the mass splitting is large, as in the singlet-doublet model with $M_1 \ll \mu$ as discussed above.

The prospects for discovering individual electroweak multiplets, whose couplings are entirely fixed by gauge invariance, were studied in~\cite{Han:2020uak}. 
The results for wino-like and higgsino-like multiplets can be mapped directly onto the limiting cases of the singlet-doublet and doublet-triplet models discussed above.
The results for wino-like and higgsino-like multiplets were quite similar. A mono-muon missing-mass search---which takes advantage of the fixed kinematics of $\mu^+\mu^-$ collisions---is projected to have a $2\sigma$ reach between 1.3 and $1.7~\textrm{TeV}$ at a 10 TeV muon collider. A combined mono-photon and disappearing track search could potentially extend this reach up to $3.1 (4.1)~\textrm{TeV}$ in the higgsino-like (wino-like) scenario. 
Note that when moving away from the pure wino- or higgsino-like limits, the mass splitting can be in the intermediate regime, inaccessible to either direct searches or searches for disappearing tracks. See also~\cite{Capdevilla:2021fmj, Capdevilla:2024bwt} for the prospects of wino- and higgsino-like dark matter in disappearing or soft-tracks.

In the rest of this paper, we will focus on parameter space in which all the charginos are kinematically accessible at a 10 TeV muon collider.
For the singlet-doublet model, the mass splitting between the chargino and the lightest neutralino is nearly always large enough such that $\psi^\pm \to \chi_1^0\, W^\pm$ is allowed. As a result, it may be feasible to extract a measurement of both the chargino and lightest neutralino masses, based on the endpoints of the $W$ energy spectrum, though a more careful analysis of the details of this measurement is necessary. See~\cite{Homiller:2022iax} for a similar study in the context of slepton decays to neutralino final states. 

The situation in the doublet-triplet model is slightly different. 
If both charginos are kinematically accessible, the lighter chargino mass can be inferred from the endpoint of the missing mass variable.
Furthermore, the mass difference between the heavier chargino and the lightest neutralino will be large enough that a prompt decay is allowed, and we would again be able to extract both masses from the endpoints of the $W$ energy spectrum. 
Thus, three separate mass measurements are possible. In terms of the model parameters, however, the mapping from the chargino masses to the Lagrangian parameters is somewhat more involved. Moreover, for light triplets, the lightest chargino and neutralino masses are degenerate, so the number of independent handles gained on the model parameters is not much different than in the singlet-doublet model.

\subsection{Indirect Constraints: Oblique Parameters and Higgs Decays}

Aside from direct searches, new electroweak fermions can leave imprints in collider experiments through their radiative effects on electroweak and Higgs precision observables. 

We first consider the effects on the oblique parameters, $S$, $T$ and $U$~\cite{Peskin:1990zt, Peskin:1991sw}.
These parameters can be readily computed in terms of the singlet-doublet and doublet-triplet model parameters in terms of the one-loop self-energies of the electroweak gauge bosons using standard formulae~\cite{Fraser:2020dpy}.

\begin{figure}[t]
\centering
\includegraphics[width=8cm]{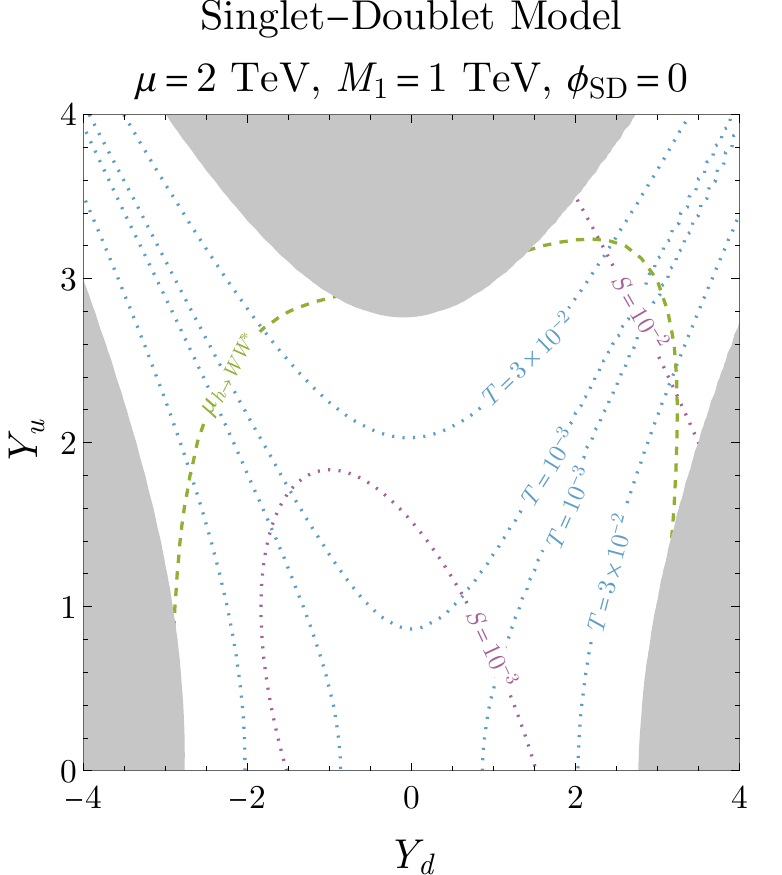}\quad
\includegraphics[width=8cm]{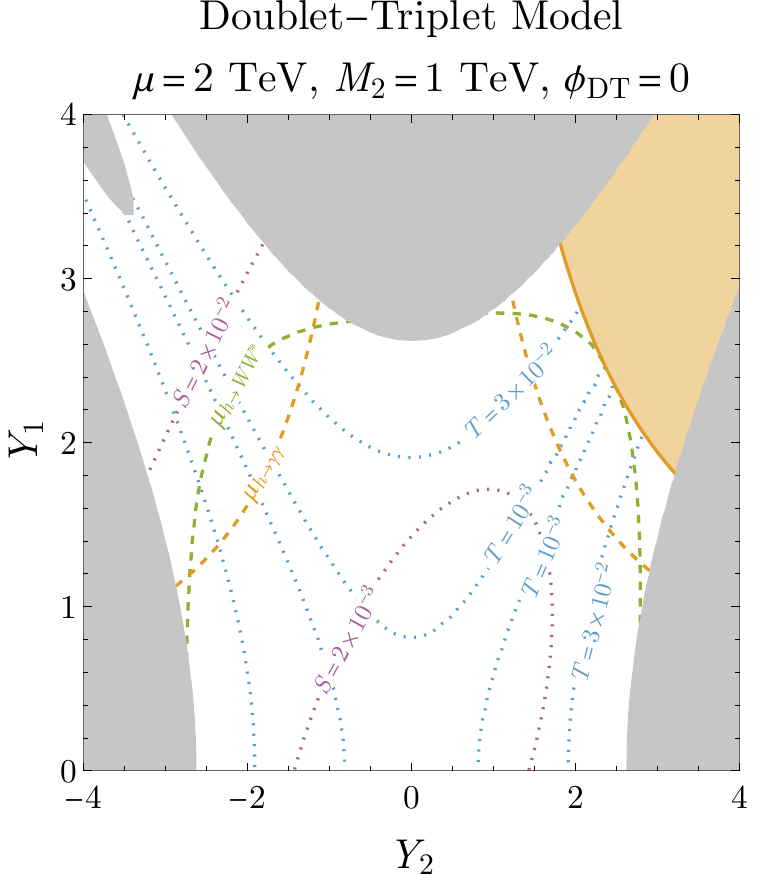}
\caption{
Indirect constraints on the singlet-doublet (left) and doublet-triplet (right) models as a function of the two independent Yukawa couplings.
The purple and blue contours show values of the oblique paramters $S$ and $T$, while the shaded gray region is ruled out by an electroweak fit to current data. The orange region (dashed curve) shows the current limit (future muon collider projection) from the $h \to \gamma\gamma$ signal strength. The dashed green line shows a similar projection for $h \to WW^*$ decays.
}\label{fig:indirect_summary1}
\end{figure}

Contours of the $S$ and $T$ parameters are shown in Fig.~\ref{fig:indirect_summary1} for the singlet-doublet (left) and doublet-triplet (right) models as a function of the two Yukawa couplings. In each case we have fixed $\mu = 2~\textrm{TeV}$ and the relevant Majorana mass $M_1$ or $M_2$ to $1$~TeV. 
The custodial symmetric limits, $Y_u = \minus Y_d$ and $Y_1 = \minus Y_2$ are apparent in the contours of the $T$ parameter; we observe that the opposite limit also leads to $T = 0$. The $S$ parameter, by contrast, vanishes only when both Yukawas are taken to zero. 
In the same plots, we show in grey the constraints from a fit to the current values of the oblique parameters based on the electroweak fit to LEP, Tevatron and LHC data performed in~\cite{ParticleDataGroup:2024cfk}. It's clear that for the $\sim $ TeV scale masses we are considering, the electroweak fit is not terribly constraining.

\begin{figure}[t!]
\centering
\includegraphics[width=8cm]{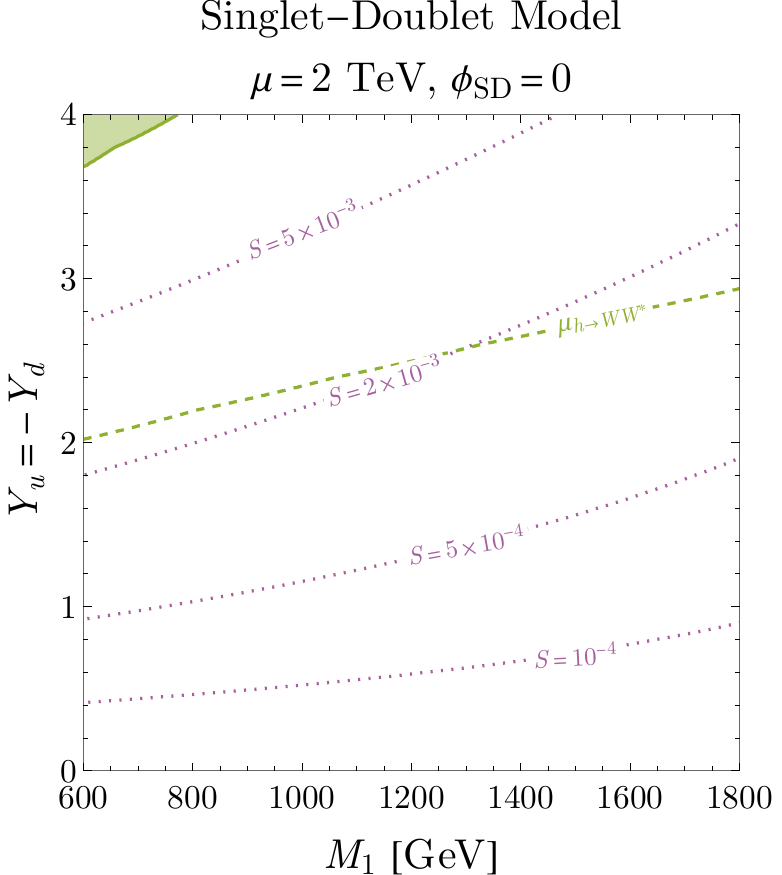}\quad
\includegraphics[width=8cm]{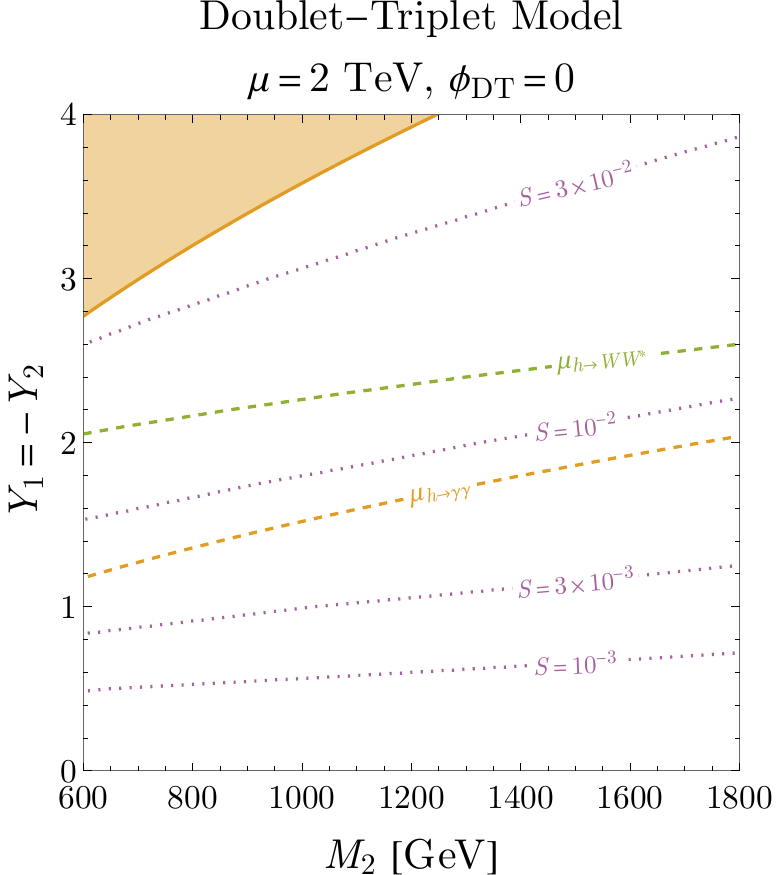}\\[1em]
\includegraphics[width=8cm]{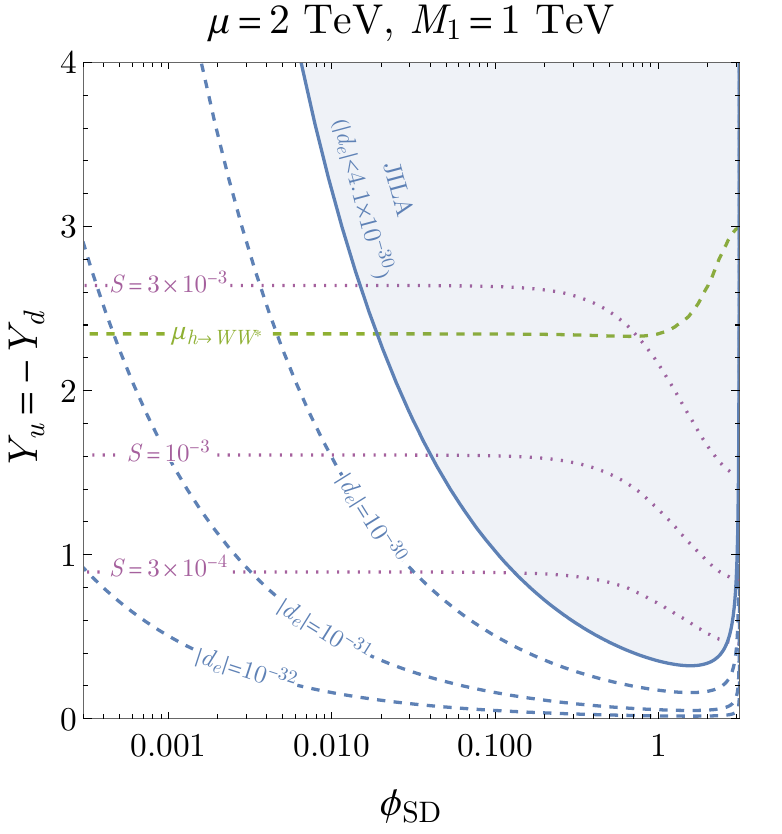}\quad
\includegraphics[width=8cm]{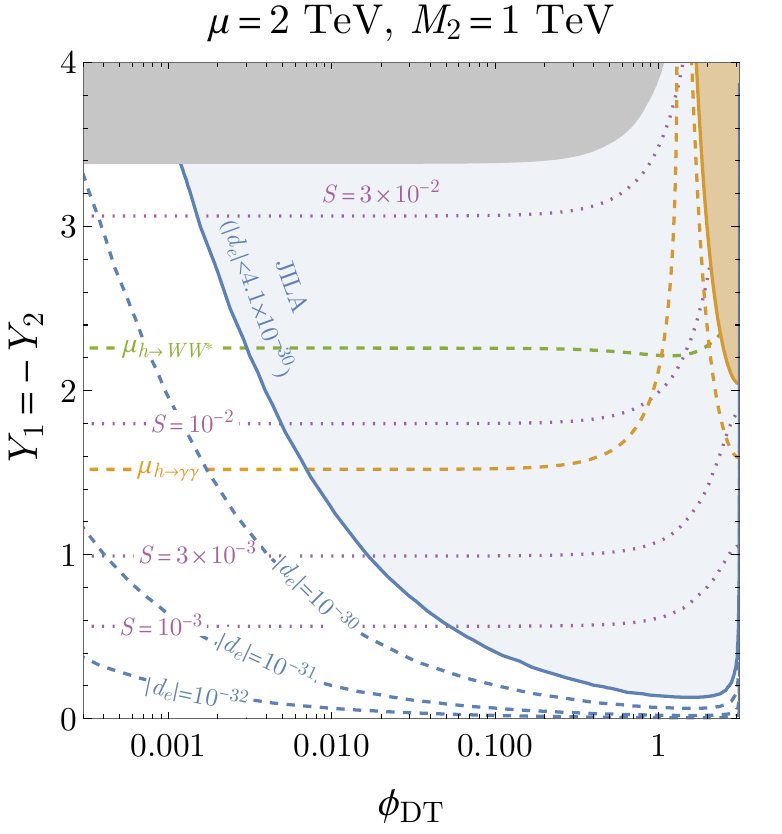}
\caption{
The same as Fig.~\ref{fig:indirect_summary1}, but in the Yukawa vs. singlet/triplet mass plane in the top row, and the Yukawa vs. CP-violating phase plane in the bottom row. In the lower panels we also show contours of constant electron EDM values, along with the current constraint, in blue.
}\label{fig:indirect_summary2}
\end{figure}

In Fig.~\ref{fig:indirect_summary2}, we show contours of the $S$ parameter as a function of other parameters of the model, taking the custodial symmetric limits for simplicity. Aside from the small region in the lower-right panel, the constraints from the electroweak fit do not appear.

\medskip

We turn next to the decays of the Higgs boson. 
In the doublet-triplet model, chargino loops contribute to the radiative $h \to \gamma\gamma$ and $h \to Z\gamma$ decays and compete with the one-loop SM contributions. The $h \to \gamma\gamma$ decay, in particular, is measured precisely by the LHC experiments, and can set important constraints, depending on the chargino masses. 

We compute the one-loop result for $h \to \gamma \gamma$ in the doublet-triplet model explicitly, including the interference with one-loop Standard Model diagrams, and compare the ratio to the SM to the combined result for the signal strength. 
Combining the ATLAS~\cite{ATLAS:2022tnm} and CMS~\cite{CMS:2021kom} constraints, both at 13~TeV, with 139 and 137$~\textrm{fb}^{-1}$ integrated luminosity respectively, we find $\mu_{\gamma\gamma} = 1.10 \pm 0.06$.
The resulting bound appears as the shaded orange  region in the right panels of Figs.~\ref{fig:indirect_summary1} and \ref{fig:indirect_summary2}. Large values of the Yukawas (especially at smaller masses) are excluded. The region $Y_1 \gg 0, Y_2 \ll 0$ is not forbidden due to the upward fluctuation in $\mu_{\gamma\gamma}$ apparent in the LHC data.

Evidence of the $h \to Z \gamma$ decay mode was only recently observed by an ATLAS and CMS combination, which found $\mu_{Z\gamma} = 2.2 \pm 0.7$~\cite{ATLAS:2023yqk}. 
The High-Luminosity phase of the LHC is projected to improve this measurement to $\mu_{Z\gamma} = 1.00 \pm 0.23$~\cite{Cepeda:2019klc}. 
We computed this contribution explicitly as well, but found that while the deviations from the Standard Model can be quite large, the precision available at the LHC (and future colliders) makes this decay mode far less constraining than $h \to \gamma\gamma$, with a similar dependence on the model parameters. As such, we will not consider it any further.

In both the singlet-doublet and doublet-triplet models, there are also one-loop corrections to the $h \to WW^*$ and $h \to ZZ^*$ decays due to the new fermions.
In this case, the loop corrections are competing with a tree-level Standard Model decay, and as a result, these corrections will be less constraining in the doublet-triplet model than the radiative decays discussed above. For the singlet-doublet model, however, these may still be one of the primary constraints.
A combination of the most recent ATLAS and CMS results~\cite{ATLAS:2016neq, ATLAS:2020rej, CMS:2022dwd} yields $\mu_{WW^*} = 1.00 \pm 0.08$ and $\mu_{ZZ*} = 1.02 \pm 0.08$.

To estimate the size of these corrections in the singlet-doublet and doublet-triplet models, we compute the full one-loop amplitudes for the three body decays $h \to W^\pm \ell\nu$ and $h \to Z\nu\nu$, neglecting the loop corrections on the $Z$ and $W$ couplings to SM fermions. This is done with numerical implementation of the renormalized model discussed in detail in Sec.~\ref{subsec:one_loop_calc}. The only visible constraint (in green) from the current measurements appears for very large values of the Yukawas in the upper-left panel of Fig.~\ref{fig:indirect_summary2}. 

\medskip 

In general, measurements of the Higgs couplings would be significantly improved at a high-energy muon collider~\cite{AlAli:2021let, Forslund:2022xjq, Forslund:2023reu}. A detailed study in~\cite{Forslund:2022xjq}, which included all background processes and detector effects via a \textsc{Delphes} fast simulation found that a 3~TeV muon collider with $1~\textrm{ab}^{-1}$ integrated luminosity could measure the $\gamma\gamma$ and $Z\gamma$ branching ratios with $6.1\%$ and $47\%$ precision, combining the $W^+W^-$ and $ZZ$ fusion production modes. At a 10~TeV collider, with $10~\textrm{ab}^{-1}$, these would be improved to $1.6\%$ and $13\%$, respectively. The precision on the $WW^*$ and $ZZ^*$ signal strengths would also be improved, to $0.42\%$ and $3.2\%$, respectively.
The $\gamma\gamma$ and $WW^*$ projected signal strengths are the most constraining, and we show them as dashed orange and green lines respectively in Figs.~\ref{fig:indirect_summary1} and \ref{fig:indirect_summary2}.

\section{One-Loop Barr--Zee Diagrams in Vector-Boson Scattering}
\label{sec:vbf}

The results reviewed in the prior section make it clear that a muon collider would have impressive capability to discover TeV scale electroweak physics that may be presaged by EDM experiments.
However, we will now show that an even more direct connection is possible: besides discovering the new states, a muon collider can test processes that are directly related to the electric dipole moment at a completely different energy scale.

The essential point, as illustrated in~\cite{AlAli:2021let}, is that the two-loop Barr--Zee diagrams are closely related to vector-boson scattering processes at a high-energy lepton collider.
Indeed, if the electron line is amputated from the diagrams in Fig.~\ref{fig:edm_diagrams}, we immediately obtain one-loop corrections to four-point amplitudes such as $\gamma\gamma \to hh$ and $WW \to h\gamma$. The additional Higgs states arise from the mass insertions on the internal fermion lines, which come (in part) from the Higgs vacuum expectation value.

While this connection relates a number of potentially interesting scattering processes, some are more amenable to measurement at a muon collider experiment than others. 
The $\gamma\gamma \to hh$ process, for instance, would appear promising---especially given the significant probability for a high-energy muon to radiate a collinear photon. However, because this process does not arise at tree-level in the SM, the leading contribution from the new particles is effectively two-loop order (in the cross section), and we find that it is too small to be measured with any precision (see also~\cite{Chiesa:2021qpr}).

We will thus confine our attention to processes which interfere with a tree-level process in the SM, modifying the rate at one-loop order. As our goal is to observe effects that are correlated with the Barr--Zee diagrams and the CP-violation, we want at least one Higgs in the final state. Of the remaining options, we choose processes with an initial state photon or $W$, due to their larger PDFs in the muon compared to the $Z$.
These considerations leave us with three candidate processes: $W^+W^- \to hh$, 
$W^+W^- \to h\gamma$ and $W^{\pm} \gamma \to W^{\pm} h$. The latter two processes are related by crossing. 

\begin{figure}[t]
\centering
\includegraphics[width=4cm]{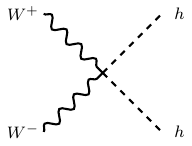}\qquad
\includegraphics[width=4cm]{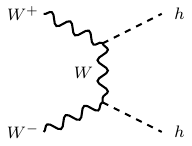}\qquad
\includegraphics[width=4cm]{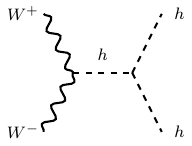}\\[0.25cm]
\includegraphics[width=4cm]{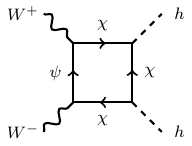}\qquad
\includegraphics[width=4cm]{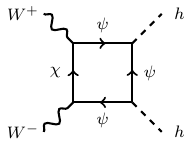}\qquad
\includegraphics[width=4cm]{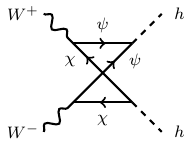}
\caption{Top row: tree-level diagrams contributing to $W^+W^- \to hh$ production relevant for a high-energy muon collider.
Bottom row: One-loop, Barr--Zee-like diagrams involving neutralinos ($\chi$) and charginos ($\psi$) that affect $W^+W^- \to hh$ production. Only the first of these diagrams exists in the singlet-doublet model, as the Higgs does not couple to the charginos. 
In both rows, additional diagrams related by interchanging identical particles are not shown.}
\label{fig:diagrams_ww_hh}
\end{figure}

\begin{figure}
\centering
\begin{minipage}{0.48\linewidth}
\centering
\includegraphics[width=4cm]{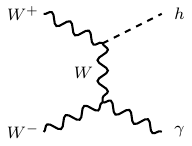}\\[0.125cm]
\includegraphics[width=4cm]{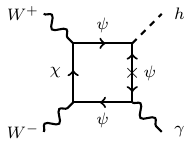}~
\includegraphics[width=4cm]{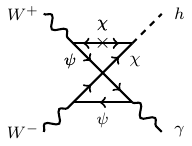}
\end{minipage}
\quad
\begin{minipage}{0.48\linewidth}
\centering
\includegraphics[width=4cm]{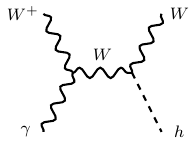}\\[0.125cm]
\includegraphics[width=4cm]{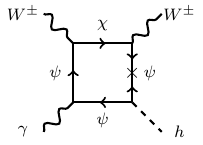}~
\includegraphics[width=4cm]{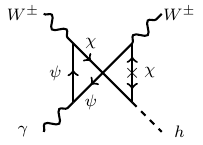}
\end{minipage}
\caption{Left: Representative diagrams leading to $W^+W^- \to h\gamma$ production at a muon collider. The top row shows the tree-level diagram, while the bottom shows the one-loop, Barr--Zee-like diagrams mediated by neutralinos ($\chi$) and charginos ($\psi$).
Note that the first of these loop diagrams does not exist in the singlet-doublet model, as the Higgs only couples to the neutralinos. 
Right: The same as the left, but for the $W^{\pm} \gamma \to W^{\pm} h$ production process, related by crossing symmetry.}
\label{fig:diagrams_ww_ha}
\end{figure}

In Fig.~\ref{fig:diagrams_ww_hh}, we show representative Feynman diagrams for the $W^+W^- \to hh$ process at tree-level (top) and one-loop (bottom). In the bottom row, we focus only on the box topologies involving the new heavy fermions.
Note that, in the singlet-doublet model, only the first of these three loop diagrams appears, as the neutral Higgs boson does not couple to the charginos.
In Fig.~\ref{fig:diagrams_ww_ha} we show the related processes $W^+W^- \to h\gamma$ (left) and $W^\pm\gamma \to W^\pm h$ (right). Again, in both cases, the first loop diagram vanishes in the singlet-doublet model. 
All three of these processes contain a number of additional diagrams at one loop, including self-energy diagrams, vertex corrections, and additional box topologies related by interchanging identical external states.
Note that, of these three processes, only $W^+W^- \to hh$ contains a four-point contact interaction at tree-level, leading to a relatively enhanced cross-section at large $\sqrt{\hat{s}}$. 

While all of these loop diagrams modify the SM process, several of them are special: those which have a discontinuity related to a unitarity cut. As we will discuss in more detail below, this discontinuity leads to a distinct kinematic feature, which makes it plausible to distinguish the loop-effects from the new states at a collider.

\subsection{Calculation of the One-Loop Effects}
\label{subsec:one_loop_calc}

We now turn to the calculation of these one-loop effects on our processes of interest. 
In general, loops involving charginos and neutralinos lead to divergences, and the theory must be renormalized.\footnote{
In fact, while the box diagrams for $W^+W^-\to hh$ are divergent, the $W^+W^- \to h\gamma$ box diagrams are actually finite: there is no operator in the SM which can absorb such a divergence. There are, however, divergent propagator and vertex corrections involved in $W^+W^- \to h\gamma$, which must be included for a reliable estimate of the one-loop effects from the new states.}
In particular, the loop diagrams in Fig.~\ref{fig:diagrams_ww_hh} lead to divergent contributions to the $W^+W^-hh$ vertex, and they must be dealt with via wave-function renormalization of the Higgs field. 
Fortunately, since we are only interested in processes involving the electroweak gauge bosons and the Higgs, and not processes with new particles as external states, we need only renormalize the electroweak sector of the Standard Model.

To renormalize the theory, and calculate the one-loop effects, we utilize the suite of tools in the \textsc{FeynArts}, \textsc{FormCalc} and \textsc{LoopTools}~\cite{Hahn:1998yk, Hahn:2000kx} packages.
The Standard Model is implemented in \textsc{FeynArts} along with the complete set of counterterms necessary for one-loop calculations in the on-shell scheme~\cite{Aoki:1980hh, Aoki:1980ix, Aoki:1982ed, Denner:1991kt}. 
We supplement this with an implementation of the new interactions involving the charginos and neutralinos in the form of a \textsc{FeynArts} model. We perform calculations in Feynman Gauge, using the on-shell scheme with $M_W$, $M_Z$ and $G_F$ chosen as the electroweak input parameters.

We validated our implementation of the model by explicitly checking that all two- and three-point Green's functions were finite after substituting in the renormalization conditions. We also directly reproduced all the calculations in~\cite{Dedes:2014hga} and \cite{Freitas:2015hsa} in the custodial-symmetric limit. 

With the renormalized amplitudes computed, we can directly compute the partonic cross sections for all our processes of interest using the numerical routines implemented in \textsc{FormCalc}.
These partonic cross sections can be convolved with the PDFs of the muon to obtain the inclusive cross sections for these processes at the collider energy of interest~\cite{Costantini:2020stv, AlAli:2021let, Ruiz:2021tdt}.
We use the public release of the muon PDFs in LePDF~\cite{Garosi:2023bvq}, which include the leading-log evolution along with all Sudakov double log, polarization, and electroweak symmetry breaking effects. We evaluate the PDFs at the scale $\sqrt{\hat{s}}/2$ for all our numerical results.

\subsection{Discontinuities in VBF Processes}

Before moving on to study the projections for measuring these processes at a high-energy muon collider, it is worthwhile to examine the properties of the loop amplitudes we have computed in more detail. As we will now show, the optical theorem leads to an interesting kinematic effect in some of these processes, which would be a smoking-gun signature of the particular Barr--Zee diagrams of interest.

As is well-known, unitarity of the $S$-matrix results in a relationship between the imaginary part of a scattering amplitude and the amplitudes for scattering of the incoming and outgoing states into all possible intermediate particles:
\begin{equation}
\label{eq:optical_theorem}
2\, \Im\, \mathcal{M}(i \to f)
= \sum_X \dd\Pi_X (2\pi)^4 \delta^4(p_i-p_X)\mathcal{M}(i\rightarrow X)\mathcal{M}^*(f\rightarrow X). 
\end{equation}
The imaginary part of loop diagrams is thus fixed by the so-called ``cutting rules''~\cite{Cutkosky:1960sp}, summing over all the ways that the loop diagrams can be cut (bringing intermediate particles on-shell) without violating momentum conservation.

In all our computations, we keep only the leading interference term between the one-loop and tree-level amplitudes. As the tree-level amplitude is manifestly real, this would seem to not depend on the imaginary part of the one-loop amplitudes that is governed by the optical theorem. 
In fact, the real and imaginary parts of the loop amplitudes are related by dispersion relations that follow from unitarity of the $S$-matrix~\cite{Remiddi:1981hn}.
As a result, we expect a kinematic feature to appear in the real part of the amplitude at intermediate particle thresholds as well, and hence in the leading corrections to the amplitude-squared. We will demonstrate this explicitly in examples.

\begin{figure}[t]
\centering
\includegraphics[width=8cm]{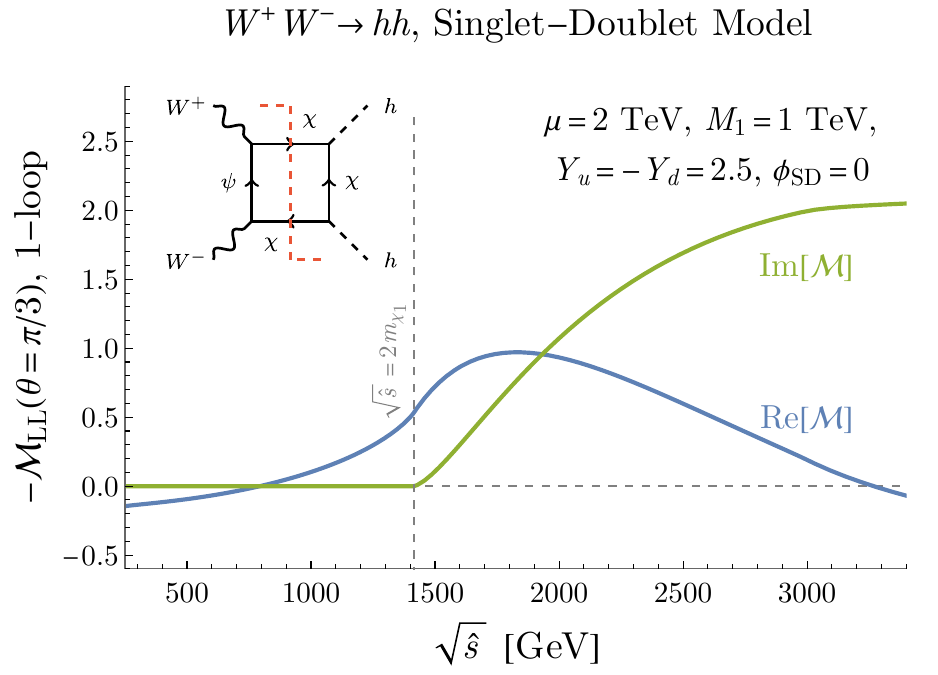}\quad
\includegraphics[width=8cm]{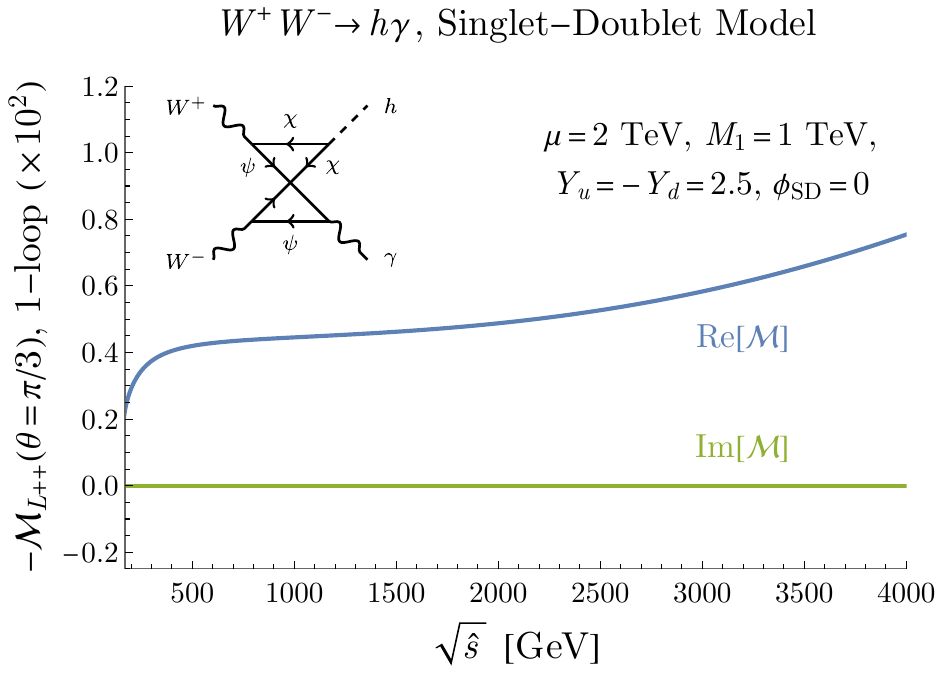}\\[0.25cm]
\includegraphics[width=8cm]{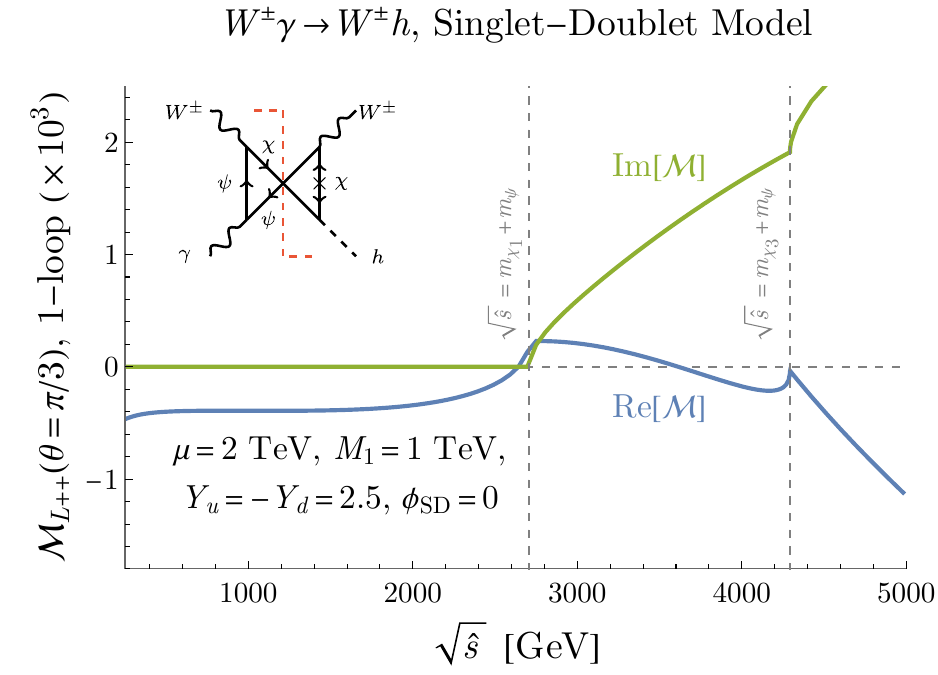}\quad
\includegraphics[width=8cm]{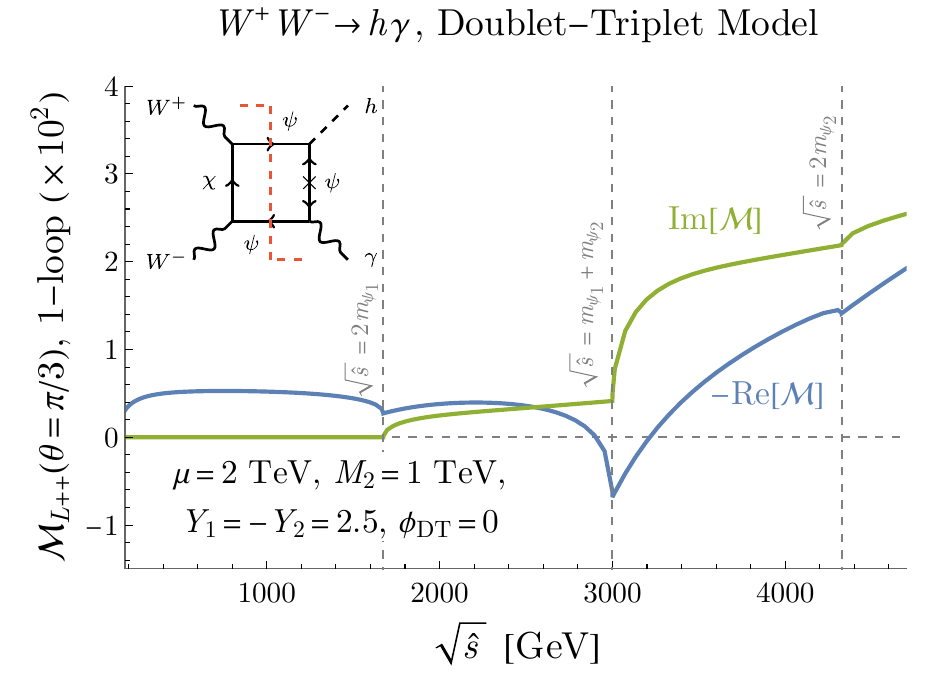}\quad
\caption{In the left column, we show examples of two diagrams with unitarity cuts, leading to discontinuities at $\sqrt{\hat{s}} = 2m_{\chi_1}$ (top) and $\sqrt{\hat{s}} = m_{\chi_i} + m_{\psi}$ (bottom). We show the results for the singlet-doublet model, but similar results exist for the doublet-triplet model as well. The process where the models show characteristically different behavior is $W^+W^-\to h\gamma$, which is shown in the right column. In the singlet-doublet model, the only diagram topology that exists does not admit a viable unitarity cut and as a result, the imaginary part is always vanishing and the real part is smooth (top). The doublet-triplet model admits diagrams with unitarity cuts at $\sqrt{\hat{s}} = m_{\psi_{i}} + m_{\psi_{j}}$ (bottom). In all these diagrams, the diagrams admitting the cuts are shown and the corresponding mass thresholds are indicated by dashed vertical lines. The discontinuities at these thresholds due to the additional imaginary part dictated by the optical theorem are clear.
}\label{fig:unitarity_cuts}
\end{figure}

Consider first $W^+W^- \to hh$ in the singlet-doublet model. In this case, only the first diagram in the bottom row of Fig.~\ref{fig:diagrams_ww_hh} appears, which has a unitarity cut at $m_{\chi_i} + m_{\chi_j}$. 
The real and imaginary parts of the total loop amplitude for this process are shown in the upper-left panel of Fig.~\ref{fig:unitarity_cuts}. The example parameter point has $m_{\chi_1} = 707~\textrm{GeV}$, and we plot only the dominant amplitude with longitudinally polarized $W$s, with the scattering angle arbitrarily fixed to $\theta = \pi/3$. 
The discontinuity in the imaginary part of the amplitude at $2m_{\chi_1}$ is readily apparent, as is the broad feature in the real part which turns on at the same point.

In the doublet-triplet model, the second and third loop diagrams in Fig.~\ref{fig:diagrams_ww_hh} contribute as well. In this case, the lightest chargino is (nearly) degenerate with the lightest neutralino, so the general features of the loop amplitude are relatively unchanged. 
Note, however, that the third loop diagram in Fig.~\ref{fig:diagrams_ww_hh} does not contain a unitarity cut: due to its topology, there is no way to consistently set two intermediate states on-shell while conserving momentum.

Turning to the processes involving a photon, we consider first $W^+W^- \to h\gamma$. In the singlet-doublet model, only the second diagram in the bottom row of Fig.~\ref{fig:diagrams_ww_ha} exists, which has no unitarity cut. This is confirmed by the top-right panel of Fig.~\ref{fig:unitarity_cuts}, which has no discernible kinematic features.

In $W^\pm\gamma \to W^\pm h$ on the other hand, the crossing symmetry changes the topology of the digram such that the corresponding bottom-right loop diagram in Fig.~\ref{fig:diagrams_ww_ha} {\em does} contain a unitarity cut at $m_{\chi_i} + m_{\psi_j}$. This is apparent in the lower-left panel of Fig.~\ref{fig:unitarity_cuts}, where we see sharp discontinuities at $\sqrt{\hat{s}} = 2700$ and $4300~\textrm{GeV}$, corresponding to $m_{\chi_1} + m_{\psi}$ and $m_{\chi_3} + m_{\psi}$.
Note that there is no feature at $m_{\chi_2} + m_{\psi}$ because the Higgs coupling to $\chi_2$ vanishes in the custodial symmetric limit.
While the features in the real part of the amplitude are apparent, they are not as broad or as large in magnitude as the feature in $W^+W^- \to hh$---this has implications for the feasibility of discovering these features.

Finally, if we consider $W^+ W^- \to h\gamma$ in the doublet-triplet model, the first diagram in the bottom row of Fig.~\ref{fig:diagrams_ww_ha} exists. The resulting features at $2m_{\psi_1}$, $m_{\psi_1} + m_{\psi_2}$ and $2 m_{\psi_2}$ are apparent in the bottom-right panel of Fig.~\ref{fig:unitarity_cuts}.
As with $W^\pm \gamma \to W^\pm h$ production, the features in the real part of the amplitude are relatively narrow.
In both these cases, the imaginary part of the one-loop amplitude has a much more significant feature, which may lead to a more noticeable effect in the tails of these distributions if the loop-squared terms are kept. Our strictly one-loop order analysis is likely rather conservative.

To summarize, the one-loop corrections from new electroweak fermions lead to distinct kinematic features in the amplitudes for vector-boson scattering processes. 
In the next section, we will explore the extent to which these features are visible at a high-energy muon collider.
Of the processes discussed above, $W^+W^- \to hh$ appears the most promising. This is in part because the tree-level amplitude contains a contact interaction, which leads to a relative enhancement at large $\sqrt{\hat{s}}$ where the feature is most prominent. The broader nature of the feature (as illustrated in Fig.~\ref{fig:unitarity_cuts}) also makes it more amenable to discovery. While the parton distribution functions for (longitudinally polarized) $W$ bosons are smaller than those for photons in a muon, the cross-section for $W^+W^-$ scattering is still sizable. Finally, in contrast to $W^+W^- \to h\gamma$ and $W^\pm\gamma \to W^\pm h$, the one-loop corrections in $W^+W^- \to hh$ include two factors of the Higgs Yukawa, which can be larger in parts of parameter space, allowing a parametric enhancement of the effect. For all these reasons, we'll focus on the $W^+W^- \to hh$ process exclusively in what follows, and leave more detailed study of the other processes to future work.

\subsection{Identifying the Discontinuity in at a High-Energy Muon Collider}
\label{subsec:analysis}

We now turn to the discovery potential for these kinematic effects at a high-energy muon collider.
In line with our premise---that we may receive the first signal of new electroweak physics from measurement of a nonzero electron EDM in the coming decade---and recalling that the direct searches discussed in Section~\ref{subsec:direct_searches} would be likely to confirm the existence of such new states, we will assume the data contains the effects of the singlet-doublet or doublet-triplet model with one of several chosen benchmark parameter points. We will first ask to what extent the resulting distribution of the Higgs-pair invariant mass in $W^+W^- \to hh$ can be distinguished from the Standard Model. We then turn to interpreting the shape of this distribution as a ``measurement'' of the underlying model parameters. 

As a first benchmark point, we consider the singlet-doublet model with $\mu = 2\,\textrm{TeV}$, $M_1 = 1\,\textrm{TeV}$, $Y_u = 2.7$, $Y_d = -2.3$ and $\phisd = 0.005$. 
This would lead to an electron EDM of $d_e = \minus 1.2 \times 10^{-30}\,e\cdot\textrm{cm}$; below the current bound but well within the reach of next-generation experiments.
This small value of the CP-violating phase could arise naturally in a more sophisticated model, e.g., if CP is a spontaneously broken symmetry. See, for instance, the discussion in~\cite{Cesarotti:2018huy}, or the explicit models in~\cite{Aloni:2021wzk}.
The chargino mass is fixed by the $\mu$ parameter, while the neutralinos have masses $708, 2001$ and $2293~\textrm{GeV}$.
At least the chargino and the lightest neutralino would thus be easily discoverable via $\mu^+\mu^- \to \psi^+\psi^-$, with the $\psi^\pm \to W^\pm \chi_1$ decay. 

\begin{figure}
\centering
\includegraphics[width=8cm]{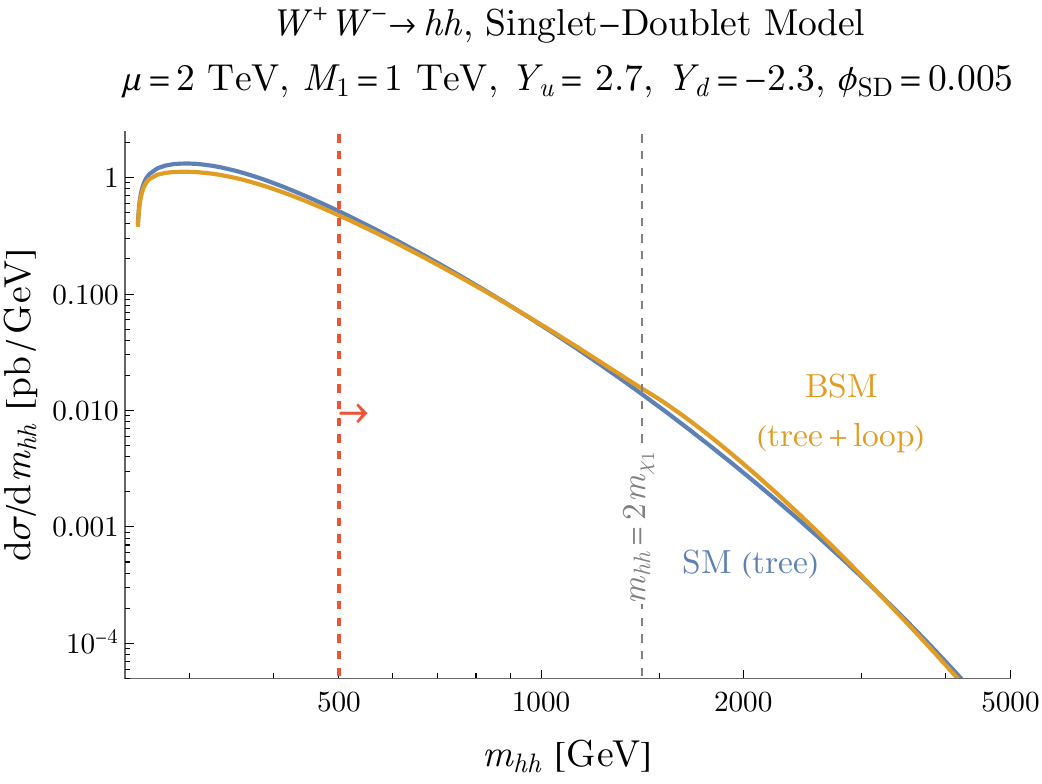}~
\includegraphics[width=8cm]{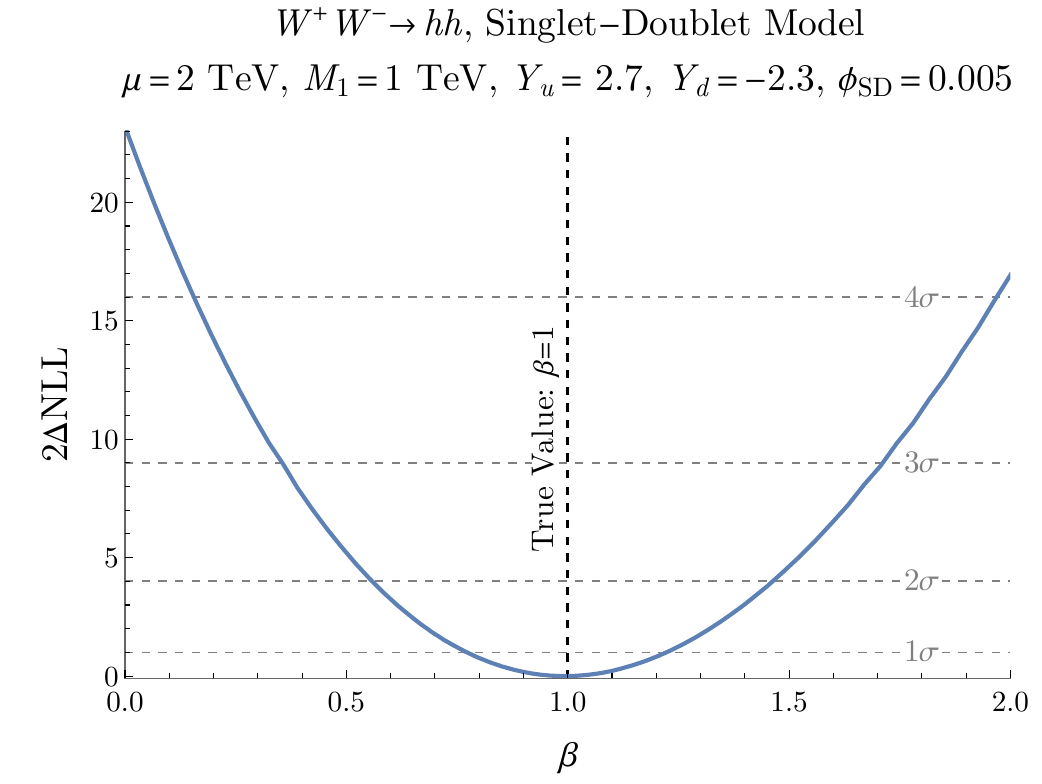}
\caption{Left: The differential cross section of $W^+W^- \to hh$. SM corresponds to the tree level SM contribution, while SM + BSM corresponds to the addition of the tree level SM contribution and the first order one-loop contribution from
the singlet-doublet model with parameters $\mu=2$ TeV, $M_1 = 1$ TeV, $Y_u=2.7, Y_d=-2.3,$ and $\phisd=0.005$. Note the kinematic feature at $m_{hh}=2m_{\chi_1}$, as predicted.
Right: $2\Delta\textrm{NLL}$ for the singlet-doublet model with parameters $\mu=2$ TeV, $M_1 = 1$ TeV,  $Y_u=2.7, Y_d=-2.3,$ and $\phisd=0.005$. $\beta=0$ corresponds to the SM distribution, while $\beta=1$ corresponds to the predicted singlet-doublet model distribution. The SM hypothesis is excluded by over $4\sigma$.
}\label{fig:special_point}
\end{figure}

The resulting differential cross section (as a function of $m_{hh} = \sqrt{\hat{s}}$) for $W^+W^- \to hh$ scattering is shown in the left panel of Fig.~\ref{fig:special_point}. This was computed using our numerical implementation of the partonic cross section, convolved with the parton luminosities of the muon, as discussed in Section~\ref{subsec:one_loop_calc}. The blue curve shows the tree-level SM distribution, while the orange curve includes the interference term from the one-loop BSM corrections. The presence of a broad feature in the orange curve starting at $m_{hh} = 2 m_{\chi_1}$ is clear.

In addition to the expected feature from the unitarity cut, there is also a clear effect from the one-loop corrections near the di-Higgs threshold.
While this is a robust, predictable effect of the new electroweak states, it is not the focus of our analysis. Moreover, the interference term between the tree-level SM amplitude and the BSM loop corrections in this region would have to compete with virtual corrections from SM particles as well, which we do not include. At higher invariant masses, these SM loop corrections will lead to a smooth correction to the slope of the distribution, which would not affect the visibility of the feature at $2 m_{\chi_1}$, but in the threshold region, the exact shape of the SM distribution would be subject to some theoretical uncertainty.

To be conservative, we will therefore focus only on the distribution with $m_{hh} > 500\,\textrm{GeV}$. We apply this cut throughout the rest of our analysis. This is far enough away from the threshold region that the SM and BSM distributions are both smooth and we do not need to worry about any other threshold effects at, e.g., twice the top quark mass. We have also explicitly calculated the one-loop corrections to the SM distribution from intermediate $t$ and $b$ quarks and massive gauge bosons, and verified that their effects become small and featureless in this regime.
We should emphasize that this cut is likely overly conservative: a complete analysis, including the threshold effects alongside a precision SM calculation would likely have much greater sensitivity to the virtual effects of new electroweak states by looking for a correlated effect in the threshold region with the effect at higher masses due to the unitarity cut. This conservative choice is made to illustrate the potential for finding the feature alone, and to simplify our analysis.

To estimate the significance of the feature in the $m_{hh}$ distribution, we generate sets of toy events taken from the BSM distribution, and perform an unbinned likelihood analysis. These toy datasets are generated assuming the SM prediction for the total cross section after the $m_{hh} > 500\,\textrm{GeV}$ cut, i.e., we do not include the effect of the loop corrections on the normalization, as the actual rate is subject to the same theoretical uncertainty in the SM discussed above. The negative log-likelihood on a set of toy events $\{x_i = m_{hh,i}\}$ is then computed as
\begin{equation}
\label{eq:}
\minus\log L(\theta) = \minus\sum_{i}^n \log f(x_i; \theta),
\end{equation}
where $\theta$ represents the full set of theory and nuisance parameters and $f(x_i; \theta)$ is the likelihood of the single event $x_i$ as a function of the parameters, computed based on the normalized distributions. Note that, consistent with excluding all information about the rate, we have dropped the Poisson term in the extended log-likelihood in the above, which would carry information about the mean number of expected events as a function of the theory parameters. As with excluding the threshold region, this again should be considered a perhaps overly conservative estimate of the sensitivity. 

Finally, again to accommodate the uncertainty in the precise shape of the Standard Model distribution, we introduce an additional, floating power-law dependence in the shape of the distribution, writing the likelihood as
\begin{equation}
\tilde{f}(x; \theta, \kappa) = \mathcal{N}_\kappa \Big(\frac{x}{\sqrt{s}}\Big)^{-\kappa} \times f(x; \theta),
\end{equation}
where $f$ is the likelihood as a function of the theory parameters computed directly. Here, $\kappa$ is a new nuisance parameter intended to capture deviations in the hardness of the distribution, and $N_\kappa$ is a constant fixed by requiring $\tilde{f}$ to be normalized. 
In all our analyses, we marginalize over the value of $\kappa$, which removes sensitivity to smooth changes in the slope from the one-loop effects. This ensures that the discriminating power comes entirely from the localized kinematic feature. Several validations that this nuisance parameter has the intended effect are presented in Appendix~\ref{app:likelihood_details}. Once again, we emphasize that this should be considered a very conservative approach to the analysis---a better understanding of the SM distribution and the detector effects involved in its measurement would allow for extra discriminating power from the overall shape of the distribution as well. 

Returning to our benchmark point, we employ this likelihood procedure to determine with what significance the SM hypothesis can be rejected. It is convenient to introduce a parameter $\beta$ and define a modified likelihood as
\begin{equation}
f(x; \beta, \kappa) \equiv (1 - \beta) f_\textsc{SM}(x;\kappa) + \beta f_{\textsc{BSM}}(x; \kappa)
\end{equation}
where $f_{\textsc{SM}, \textsc{BSM}}$ are the p.d.f.s of the SM and benchmark point, respectively. This smoothly interpolates between the SM at $\beta = 0$ and the benchmark value at $\beta = 1$. We then evaluate the log-likelihood as a function of $\beta$ on 100 toy datasets sampled from the BSM distribution. The resulting difference in twice the negative log likelihood ($2\Delta\textrm{NLL}$) as a function of $\beta$ is shown in the right panel of Fig.~\ref{fig:special_point}. According to Wilks' theorem, $2\Delta\textrm{NLL}$ will asymptote to a $\chi^2$ distribution with one degree of freedom~\cite{Wilks:1938dza, Cowan:2010js}. The resulting confidence levels are indicated by the dashed horizontal lines in Fig.~\ref{fig:special_point}. We see that this benchmark point could be distinguished from the SM at over $4 \sigma$. 

\medskip

\begin{figure}
\centering
\includegraphics[width=8cm]{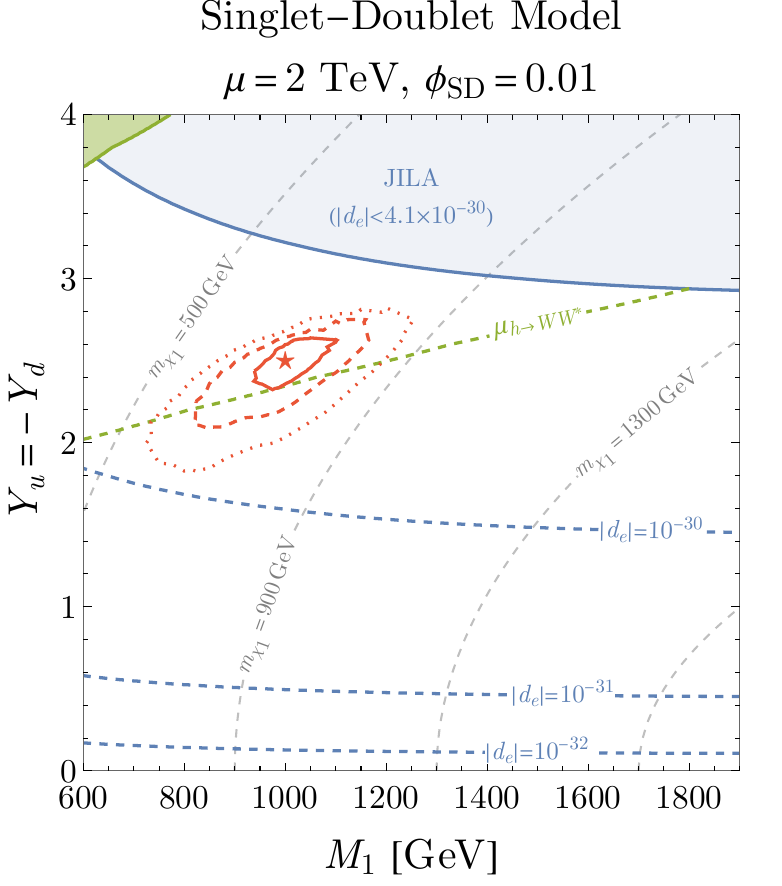}~
\includegraphics[width=8cm]{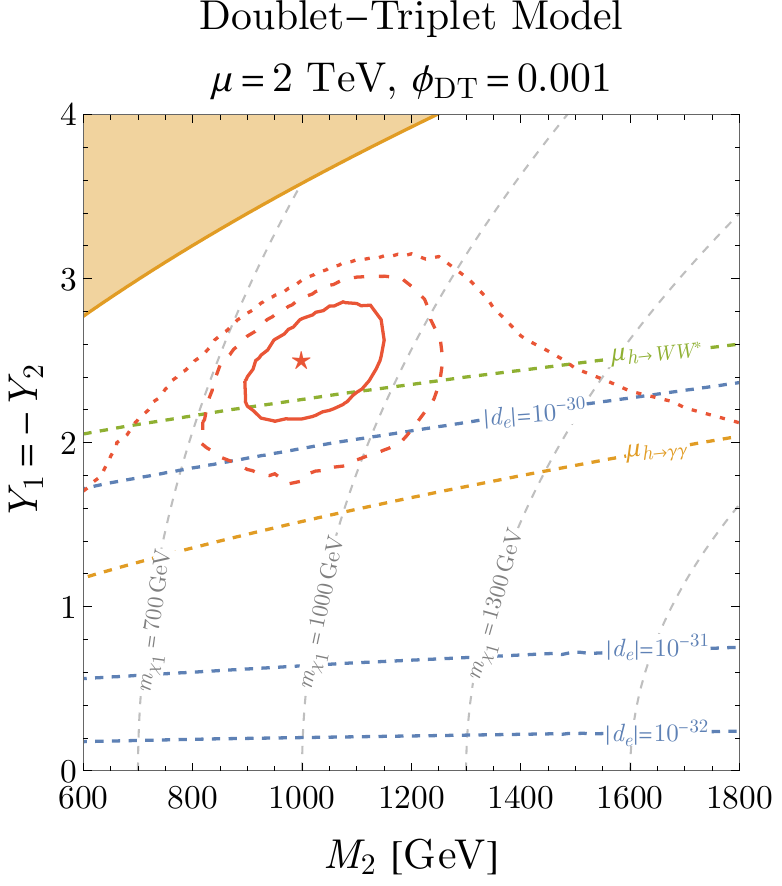}\\[1em]
\includegraphics[width=8cm]{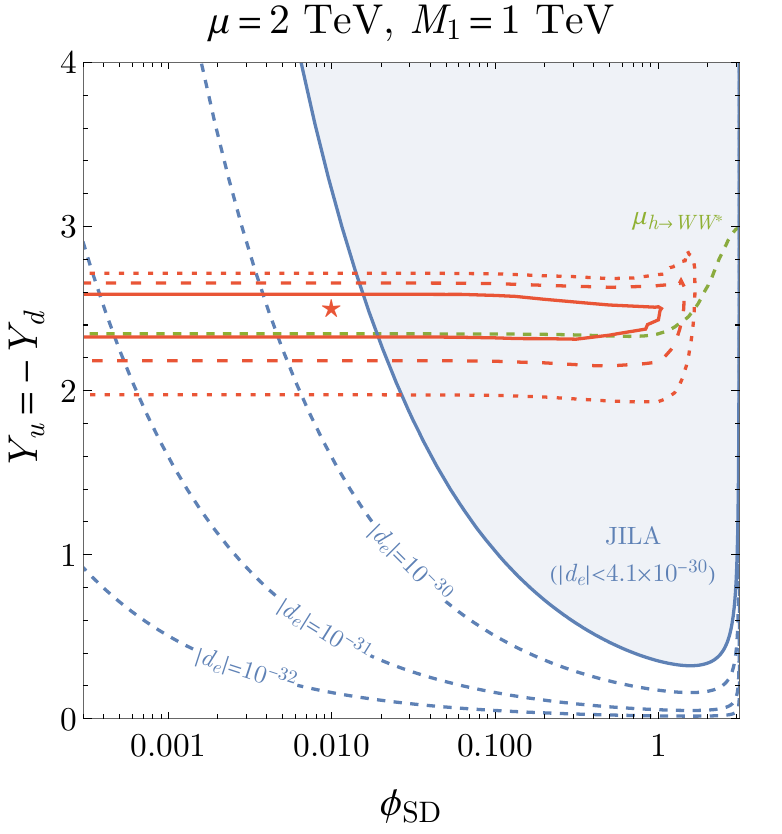}~
\includegraphics[width=8cm]{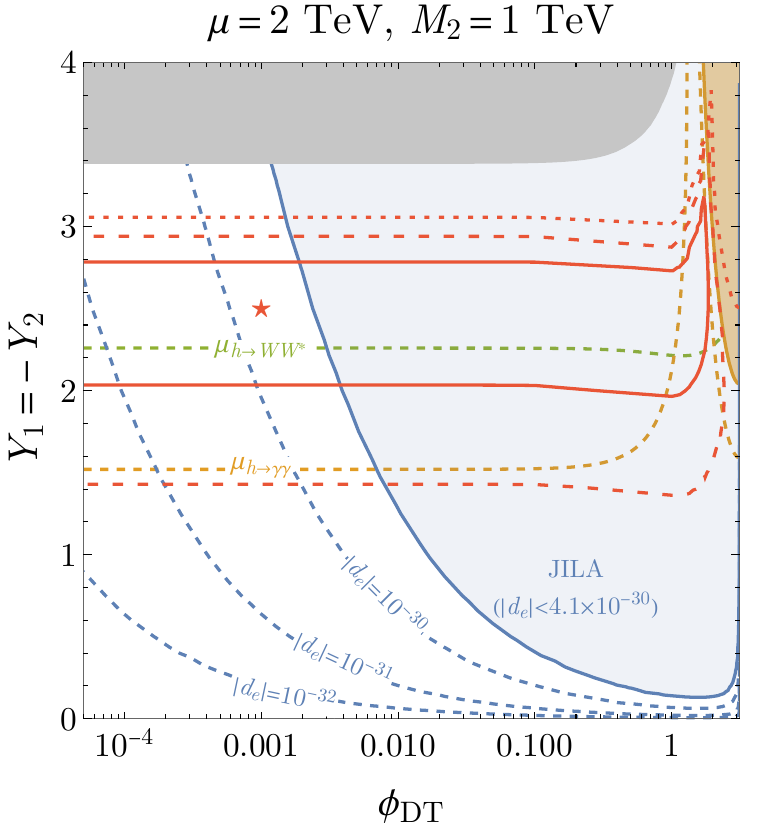}
\caption{
Contours of the $1, 2$ and $3\sigma$ preferred regions from the likelihood analysis as solid, dashed and dotted red lines in parameter planes similar to Fig.~\ref{fig:indirect_summary2}. The true value used to generate the toy datasets is indicated by the red star. The dashed gray lines show contours of constant $m_{\chi_1}$. Contours of the eEDM are shown in blue, as before, and other constraints and projections are as in Fig.~\ref{fig:indirect_summary2}.
}\label{fig:likelihood_scan_2d}
\end{figure}

We next turn to interpreting the $m_{hh}$ distribution as a measurement of the singlet-doublet or doublet-triplet model parameters. For convenience, we work in the custodial symmetric limit ($Y_u = \minus Y_d$ and $Y_1 = \minus Y_2$, respectively). 
This limit maximizes the Higgs couplings to the lightest neutralino (and chargino, in the doublet-triplet model), while setting the Higgs coupling to $\chi_2$ to zero, but does not otherwise lead to any additional enhancement or suppression of the signal. 
We further fix $\mu = 2\,\textrm{TeV}$ in each model, as a benchmark, and can then look at the measurements in the same parameter planes as illustrated in Fig.~\ref{fig:indirect_summary2}.
For the singlet-doublet model, we generate the toy datasets with $M_1 = 1~\textrm{TeV}$, $Y_u = \minus Y_d = 2.5$, and $\phisd = 0.01$. 
This leads to neutralino masses of $707, 2000$ and $2293~\textrm{GeV}$ and an EDM $d_e = \minus 2.5 \times 10^{-30}\,e\cdot\textrm{cm}$. 
We use similar parameters for the doublet-triplet model: $M_2 = 1~\textrm{TeV}$, $Y_1 = \minus Y_2 = 2.5$ but $\phidt = 10^{-3}$. 
The neutralino masses in this case are $837$, $2000$ and $2163~\textrm{GeV}$, the charginos are degenerate with $\chi_1$ and $\chi_3$, and the resulting electron EDM is $d_e = 1.7 \times 10^{-30}\,e\cdot\textrm{cm}$
In all cases, we compute the likelihood as a function of the two parameters, keeping the others fixed to the correct value.

The results are shown in Fig.~\ref{fig:likelihood_scan_2d}, alongside contours of the electron EDM. Other than the appearance of the EDM contours, due to the nonzero CP-violating phase, the other constraints in the top row are essentially the same as in Fig.~\ref{fig:indirect_summary1}. 
The red star indicates the ``true'' value of the parameters used to generate the toy datasets, and the solid, dashed and dotted red lines indicate the $1$, $2$ and $3\sigma$ confidence level regions respectively, computed the same way as above. Note that in the doublet-triplet case, the lower boundary of the dotted line does not appear, as the limit $Y_1 = \minus Y_2 \to 0$ is not excluded at $3\sigma$.

Several qualitative features of these preferred regions are immediately apparent. In all cases, we see that a reasonable measurement of the Yukawa couplings is possible, with somewhat better precision achievable in the singlet-doublet model.
This reflects the direct dependence of the ``height'' of the kinematic feature on the size of the Yukawa couplings.  
For a fixed value of the phase, the underlying parameter $M_1$ or $M_2$ can also be measured with $\mathcal{O}(100\,\textrm{GeV})$ precision via the $W^+W^- \to hh$ distribution. 
The orientation of the preferred regions in the top row of Fig.~\ref{fig:likelihood_scan_2d} roughly follows contours of constant $m_{\chi_1}$, as expected because this determines the location of the kinematic feature, but the degeneracy is largely lifted due to the aforementioned dependence on the absolute value of the Yukawas.

The information on the phase, on the other hand, is much less precise---only an $\mathcal{O}(1)$ measurement of the phase is possible, even for fixed masses and Yukawa couplings. 
This is not surprising, because for small enough values of the phase, the mass and effective coupling to the Higgs of the lightest neutralino remains relatively unchanged. For $\mathcal{O}(1)$ values of the phase, we start noticing $\mathcal{O}(10\%)$ shifts in the effective couplings and mass. This is reflected in the bottom row of Fig.~\ref{fig:likelihood_scan_2d}, where the contours remain relatively flat for small values of the phase and start to deviate when the phase becomes $\mathcal{O}(1)$.
That said, we see that the preferred regions from the $m_{hh}$ distribution measurement are nearly orthogonal to the contours of constant $d_e$. Thus, despite coming from the same Barr--Zee-like diagrams, the electron EDM and vector-boson scattering observables provide quite complementary information on the underlying theory. 

Combining the information on the mass spectrum from direct searches, a measurement of the electron EDM, and the measurement of the size and location of the feature, in principle, provides enough information to constrain all the independent parameters of these models. Measurements of the deviations in the Higgs signal strengths would provide additional input. Note that, while we worked in the custodial symmetric limit above for simplicity, a measurement of the kinematic feature in $m_{hh}$ would fix one combination of the Yukawa couplings in the more general case.

\section{Conclusions}
\label{sec:conclusion}

While we have yet to see convincing signs of beyond the Standard Model physics, the increasing sensitivity of precision probes implies that this situation could change dramatically in the near future. If a nonzero electron EDM is found, detailed scrutiny of the two-loop interpretation---involving Barr--Zee diagrams of new, electroweak particles near the TeV scale---would be mandatory.

Using two simplified models as benchmarks, we have demonstrated that a high-energy muon collider would have unique capabilities in performing such scrutiny. 
Aside from the ability to directly produce and detect the new electroweak states, we illustrated that the connection between Barr--Zee diagrams and vector-boson scattering processes leads to an indirect, complementary method for studying such models. The vector-boson scattering processes generically contain interesting kinematic features as a result of the unitarity of the $S$-matrix, and this leaves an imprint that can be detected at a high-energy machine.
Beyond providing a more direct connection between a nonzero EDM and the discovery of new electroweak states, a measurement of this feature in vector-boson scattering would also provide complementary information on the parameters of the underlying theory. 

Such a measurement is possible only in the unique environment promised by a high-energy lepton collider: the combination of large vector-boson scattering rates and an environment clean from QCD backgrounds in which Higgs pairs can be reliably identified is only possible in this context.

There are ample opportunities for further exploring the capabilities of a high-energy muon collider to discover new physics via vector-boson scattering.
This is only one example of the possibilities that a vector-boson collider presents. It would be interesting to study the possibilities of observing virtual effects in ``electroweak precision'' measurements at a muon collider more generally, particularly in comparison to future prospects at the $Z$-pole.
It would also be interesting to search for other processes which are sensitive to new sources of CP violation more directly, both at $\mu^+\mu^-$ and $\mu^+\mu^+$ colliders, the latter of which has been proposed as a nearer-term possibility based on ultra-cold muon technology using intermediate muonium, developed for the muon $g-2$ experiment at J-PARC~\cite{Hamada:2022mua}. The $\mu^+$ beams generated with this scheme can be polarized, which presents more potential opportunities for CP studies that deserve scrutiny in future work.\footnote{We thank the anonymous referee for pointing this out to us.} 
Our work, and the capabilities of VBF measurements more generally, also motivate continued advancements in the Standard Model predictions for these processes at a high-energy lepton machine. 
Aside from Standard Model loop corrections, advancements in computing the gauge boson PDFs in the muon, and matching these inclusive calculations with additional partons, are urgently needed. 
Clearly, there is lots of exciting work to be done.

\section*{Acknowledgments}

We are grateful to Matthew Forslund and Weishuang Linda Xu for useful discussions, and especially to Ayres Freitas for sharing his \textsc{FeynArts} models for the singlet-doublet and doublet-triplet models to help us check our results.
The work of SH is supported by the NSF grant PHY-2309456. The work of JL is supported by the Moore Foundation Award 8342 and Harvard University institutional funding. MR is supported in part by the DOE Grant DE-SC0013607.
The work of AP is supported in part by the US National Science Foundation Grant PHY2210533 and the Simons Foundation Grant Nos.~623940 and MPS-SIP-00010469.

\appendix

\section{Additional Details for the Likelihood Analysis}
\label{app:likelihood_details}

In this appendix, we briefly present some checks on the robustness of our likelihood procedure. Our goal is to demonstrate that fit is sensitive predominantly to the size and location of the kinematic feature, and not to smooth changes in the overall shape of the distribution, which might be mimicked by higher-order corrections in the SM or by detector effects.

\begin{figure}[t]
\centering
\includegraphics[width=8cm]{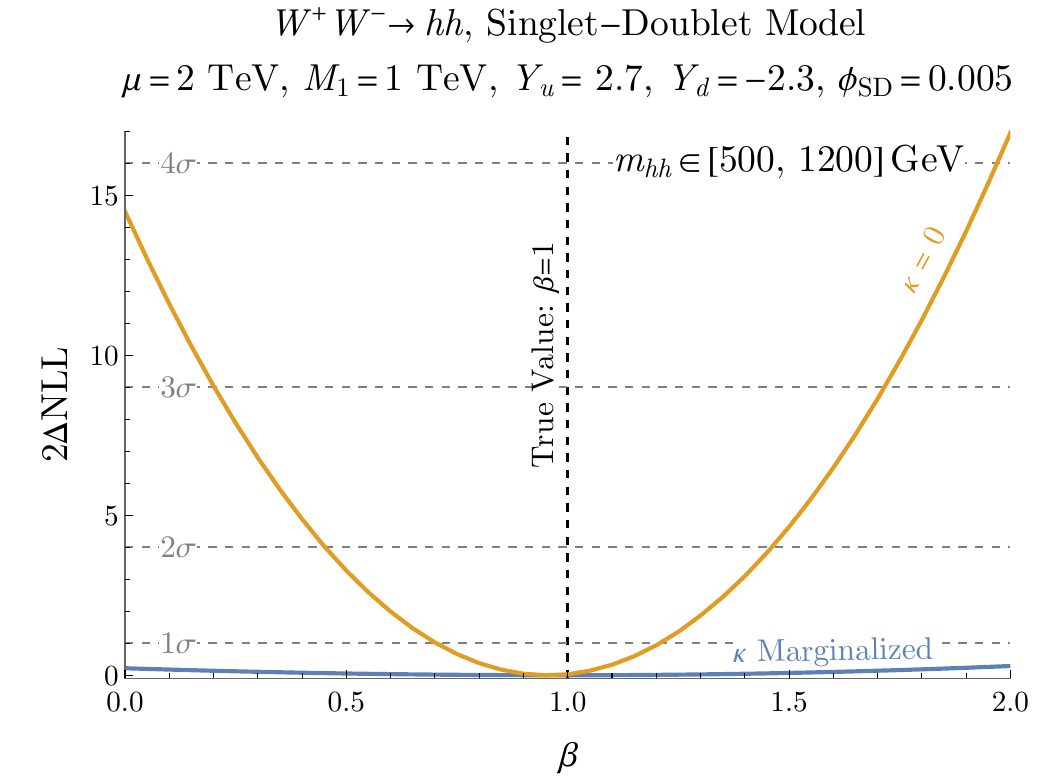}\quad
\includegraphics[width=8cm]{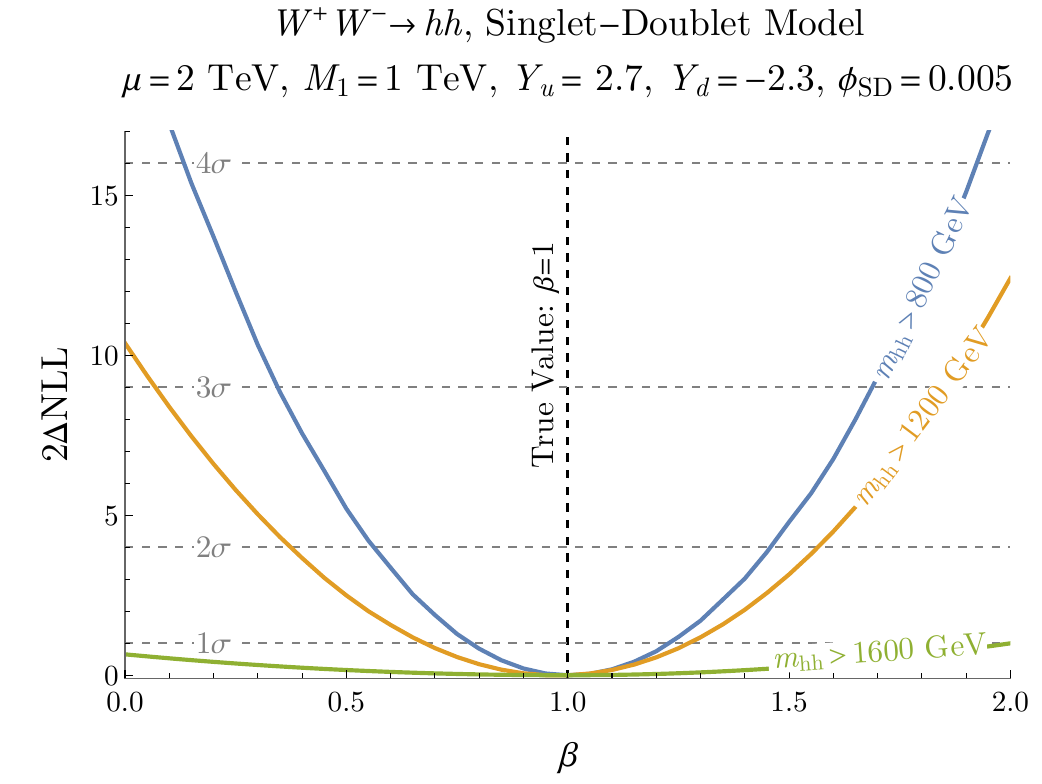}\quad
\caption{Left: Twice the negative log-likelihood when the analysis is performed only on the range $m_{hh} \in [500, 1200]\,\textrm{GeV}$, which excludes the feature. The blue (orange) curves show the procedure with (without) marginalizing over the additional power-law nuisance parameter. 
Right: The same as the left plot, but with only a lower bound on $m_{hh}$ at 800~GeV (blue), 1200~GeV (orange) and 1600~GeV (green).
}\label{fig:robustness_tests}
\end{figure}

We first perform the same analysis for the single benchmark point as in the beginning of Section~\ref{subsec:analysis}, except that we include only the events with $m_{hh}$ between $500$ and $1200~\textrm{GeV}$, i.e., events that do not include the kinematic feature of interest at $m_{hh} = 2 m_{\chi_1} = 1415\,\textrm{GeV}$. 
The results are shown in the left panel of Fig.~\ref{fig:robustness_tests}, where the blue curve shows the likelihood after marginalizing over $\kappa$ as usual, while the orange curve shows the results with $\kappa$ fixed to $0$.
Clearly, while the result is less significant than the analysis with the kinematic feature included, the change in the slope of the distribution provides some discriminating power between the SM and BSM distributions. Marginalizing over the additional nuisance parameter completely eliminates this sensitivity, as desired.

We next perform the same analysis with no upper bound on $m_{hh}$, but changing the lower bound on the invariant mass to different values. The results (all including the marginalization over $\kappa$) are shown in the right panel of Fig.~\ref{fig:robustness_tests}. We see that the discriminating power decreases modestly as the cut is moved from $800$ to $1200~\textrm{GeV}$, which reflects that sufficient statistics for $m_{hh} < 2 m_{\chi_1}$ are necessary to identify the discontinuity. Once the cut is moved past the kinematic feature to $1600~\textrm{GeV}$, the discriminating power essentially vanishes, as expected.

{\small
\bibliography{muc-edm-refs}

\providecommand{\href}[2]{#2}\begingroup\raggedright\begin{thebibliography}{100}

\bibitem{Davidson:2022jai}
S.~Davidson, B.~Echenard, R.~H. Bernstein, J.~Heeck, and D.~G. Hitlin,
  ``{Charged Lepton Flavor Violation},''
  \href{http://arxiv.org/abs/2209.00142}{{\tt arXiv:2209.00142 [hep-ex]}}.

\bibitem{Alarcon:2022ero}
R.~Alarcon {\em et al.}, ``{Electric dipole moments and the search for new
  physics},'' in {\em {Snowmass 2021}}.
\newblock 3, 2022.
\newblock \href{http://arxiv.org/abs/2203.08103}{{\tt arXiv:2203.08103
  [hep-ph]}}.

\bibitem{Roussy:2022cmp}
T.~S. Roussy {\em et al.}, ``{An improved bound on the
  electron\textquoteright{}s electric dipole moment},''
  \href{http://dx.doi.org/10.1126/science.adg4084}{{\em Science} {\bf 381}
  (2023) no.~6653, adg4084}, \href{http://arxiv.org/abs/2212.11841}{{\tt
  arXiv:2212.11841 [physics.atom-ph]}}.

\bibitem{Cesarotti:2018huy}
C.~Cesarotti, Q.~Lu, Y.~Nakai, A.~Parikh, and M.~Reece, ``{Interpreting the
  Electron EDM Constraint},''
  \href{http://dx.doi.org/10.1007/JHEP05(2019)059}{{\em JHEP} {\bf 05} (2019)
  059}, \href{http://arxiv.org/abs/1810.07736}{{\tt arXiv:1810.07736
  [hep-ph]}}.

\bibitem{Choi:1990cn}
K.~Choi and J.-y. Hong, ``{Electron electric dipole moment and Theta (QCD)},''
  \href{http://dx.doi.org/10.1016/0370-2693(91)90838-H}{{\em Phys. Lett. B}
  {\bf 259} (1991)  340--344}.

\bibitem{Ghosh:2017uqq}
D.~Ghosh and R.~Sato, ``{Lepton Electric Dipole Moment and Strong CP
  Violation},'' \href{http://dx.doi.org/10.1016/j.physletb.2017.12.052}{{\em
  Phys. Lett. B} {\bf 777} (2018)  335--339},
  \href{http://arxiv.org/abs/1709.05866}{{\tt arXiv:1709.05866 [hep-ph]}}.

\bibitem{Pospelov:2013sca}
M.~Pospelov and A.~Ritz, ``{CKM benchmarks for electron electric dipole moment
  experiments},'' \href{http://dx.doi.org/10.1103/PhysRevD.89.056006}{{\em
  Phys. Rev. D} {\bf 89} (2014) no.~5, 056006},
  \href{http://arxiv.org/abs/1311.5537}{{\tt arXiv:1311.5537 [hep-ph]}}.

\bibitem{Hoogeveen:1990cb}
F.~Hoogeveen, ``{The Standard Model Prediction for the Electric Dipole Moment
  of the Electron},''
  \href{http://dx.doi.org/10.1016/0550-3213(90)90182-D}{{\em Nucl. Phys. B}
  {\bf 341} (1990)  322--340}.

\bibitem{Pospelov:1991zt}
M.~E. Pospelov and I.~B. Khriplovich, ``{Electric dipole moment of the W boson
  and the electron in the Kobayashi-Maskawa model},'' {\em Sov. J. Nucl. Phys.}
  {\bf 53} (1991)  638--640.

\bibitem{Ema:2022yra}
Y.~Ema, T.~Gao, and M.~Pospelov, ``{Standard Model Prediction for Paramagnetic
  Electric Dipole Moments},''
  \href{http://dx.doi.org/10.1103/PhysRevLett.129.231801}{{\em Phys. Rev.
  Lett.} {\bf 129} (2022) no.~23, 231801},
  \href{http://arxiv.org/abs/2202.10524}{{\tt arXiv:2202.10524 [hep-ph]}}.

\bibitem{Flambaum:2019ejc}
V.~V. Flambaum, M.~Pospelov, A.~Ritz, and Y.~V. Stadnik, ``{Sensitivity of EDM
  experiments in paramagnetic atoms and molecules to hadronic CP violation},''
  \href{http://dx.doi.org/10.1103/PhysRevD.102.035001}{{\em Phys. Rev. D} {\bf
  102} (2020) no.~3, 035001}, \href{http://arxiv.org/abs/1912.13129}{{\tt
  arXiv:1912.13129 [hep-ph]}}.

\bibitem{Barr:1990vd}
S.~M. Barr and A.~Zee, ``{Electric Dipole Moment of the Electron and of the
  Neutron},'' \href{http://dx.doi.org/10.1103/PhysRevLett.65.21}{{\em Phys.
  Rev. Lett.} {\bf 65} (1990)  21--24}. [Erratum: Phys.Rev.Lett. 65, 2920
  (1990)].

\bibitem{AlAli:2021let}
H.~Al~Ali {\em et al.}, ``{The muon Smasher\textquoteright{}s guide},''
  \href{http://dx.doi.org/10.1088/1361-6633/ac6678}{{\em Rept. Prog. Phys.}
  {\bf 85} (2022) no.~8, 084201}, \href{http://arxiv.org/abs/2103.14043}{{\tt
  arXiv:2103.14043 [hep-ph]}}.

\bibitem{Aime:2022flm}
C.~Aime {\em et al.}, ``{Muon Collider Physics Summary},''
  \href{http://arxiv.org/abs/2203.07256}{{\tt arXiv:2203.07256 [hep-ph]}}.

\bibitem{MuonCollider:2022nsa}
{\bf Muon Collider} Collaboration, D.~Stratakis {\em et al.}, ``{A Muon
  Collider Facility for Physics Discovery},''
  \href{http://arxiv.org/abs/2203.08033}{{\tt arXiv:2203.08033
  [physics.acc-ph]}}.

\bibitem{Black:2022cth}
K.~M. Black {\em et al.}, ``{Muon Collider Forum report},''
  \href{http://dx.doi.org/10.1088/1748-0221/19/02/T02015}{{\em JINST} {\bf 19}
  (2024) no.~02, T02015}, \href{http://arxiv.org/abs/2209.01318}{{\tt
  arXiv:2209.01318 [hep-ex]}}.

\bibitem{MuonCollider:2022glg}
{\bf Muon Collider} Collaboration, S.~Jindariani {\em et al.}, ``{Promising
  Technologies and R\&D Directions for the Future Muon Collider Detectors},''
  \href{http://arxiv.org/abs/2203.07224}{{\tt arXiv:2203.07224
  [physics.ins-det]}}.

\bibitem{MuonCollider:2022ded}
{\bf Muon Collider} Collaboration, N.~Bartosik {\em et al.}, ``{Simulated
  Detector Performance at the Muon Collider},''
  \href{http://arxiv.org/abs/2203.07964}{{\tt arXiv:2203.07964 [hep-ex]}}.

\bibitem{Bartosik:2024ulr}
{\bf International Muon Collider} Collaboration, N.~Bartosik, D.~Calzolari,
  L.~Castelli, A.~Lechner, and D.~Lucchesi, ``{Machine-Detector interface for
  multi-TeV Muon Collider},'' \href{http://dx.doi.org/10.22323/1.449.0630}{{\em
  PoS} {\bf EPS-HEP2023} (2024)  630}.

\bibitem{Skoufaris:2024okx}
{\bf International Muon Collider} Collaboration, K.~Skoufaris, ``{Status of the
  International Muon Collider Complex Study at 10 TeV},''
  \href{http://dx.doi.org/10.22323/1.449.0629}{{\em PoS} {\bf EPS-HEP2023}
  (2024)  629}.

\bibitem{InternationalMuonCollider:2024jyv}
{\bf International Muon Collider} Collaboration, C.~Accettura {\em et al.},
  ``{Interim report for the International Muon Collider Collaboration
  (IMCC)},'' \href{http://arxiv.org/abs/2407.12450}{{\tt arXiv:2407.12450
  [physics.acc-ph]}}.

\bibitem{Chiesa:2020awd}
M.~Chiesa, F.~Maltoni, L.~Mantani, B.~Mele, F.~Piccinini, and X.~Zhao,
  ``{Measuring the quartic Higgs self-coupling at a multi-TeV muon collider},''
  \href{http://dx.doi.org/10.1007/JHEP09(2020)098}{{\em JHEP} {\bf 09} (2020)
  098}, \href{http://arxiv.org/abs/2003.13628}{{\tt arXiv:2003.13628
  [hep-ph]}}.

\bibitem{Han:2020pif}
T.~Han, D.~Liu, I.~Low, and X.~Wang, ``{Electroweak couplings of the Higgs
  boson at a multi-TeV muon collider},''
  \href{http://dx.doi.org/10.1103/PhysRevD.103.013002}{{\em Phys. Rev. D} {\bf
  103} (2021) no.~1, 013002}, \href{http://arxiv.org/abs/2008.12204}{{\tt
  arXiv:2008.12204 [hep-ph]}}.

\bibitem{Buttazzo:2020uzc}
D.~Buttazzo, R.~Franceschini, and A.~Wulzer, ``{Two Paths Towards Precision at
  a Very High Energy Lepton Collider},''
  \href{http://dx.doi.org/10.1007/JHEP05(2021)219}{{\em JHEP} {\bf 05} (2021)
  219}, \href{http://arxiv.org/abs/2012.11555}{{\tt arXiv:2012.11555
  [hep-ph]}}.

\bibitem{Forslund:2022xjq}
M.~Forslund and P.~Meade, ``{High precision higgs from high energy muon
  colliders},'' \href{http://dx.doi.org/10.1007/JHEP08(2022)185}{{\em JHEP}
  {\bf 08} (2022)  185}, \href{http://arxiv.org/abs/2203.09425}{{\tt
  arXiv:2203.09425 [hep-ph]}}.

\bibitem{Forslund:2023reu}
M.~Forslund and P.~Meade, ``{Precision Higgs width and couplings with a high
  energy muon collider},''
  \href{http://dx.doi.org/10.1007/JHEP01(2024)182}{{\em JHEP} {\bf 01} (2024)
  182}, \href{http://arxiv.org/abs/2308.02633}{{\tt arXiv:2308.02633
  [hep-ph]}}.

\bibitem{Han:2023njx}
T.~Han, D.~Liu, I.~Low, and X.~Wang, ``{Electroweak scattering at the muon
  shot},'' \href{http://dx.doi.org/10.1103/PhysRevD.110.013005}{{\em Phys. Rev.
  D} {\bf 110} (2024) no.~1, 013005},
  \href{http://arxiv.org/abs/2312.07670}{{\tt arXiv:2312.07670 [hep-ph]}}.

\bibitem{Li:2024joa}
P.~Li, Z.~Liu, and K.-F. Lyu, ``{Higgs boson width and couplings at high energy
  muon colliders with forward muon detection},''
  \href{http://dx.doi.org/10.1103/PhysRevD.109.073009}{{\em Phys. Rev. D} {\bf
  109} (2024) no.~7, 073009}, \href{http://arxiv.org/abs/2401.08756}{{\tt
  arXiv:2401.08756 [hep-ph]}}.

\bibitem{Buttazzo:2018qqp}
D.~Buttazzo, D.~Redigolo, F.~Sala, and A.~Tesi, ``{Fusing Vectors into Scalars
  at High Energy Lepton Colliders},''
  \href{http://dx.doi.org/10.1007/JHEP11(2018)144}{{\em JHEP} {\bf 11} (2018)
  144}, \href{http://arxiv.org/abs/1807.04743}{{\tt arXiv:1807.04743
  [hep-ph]}}.

\bibitem{Asadi:2021gah}
P.~Asadi, R.~Capdevilla, C.~Cesarotti, and S.~Homiller, ``{Searching for
  leptoquarks at future muon colliders},''
  \href{http://dx.doi.org/10.1007/JHEP10(2021)182}{{\em JHEP} {\bf 10} (2021)
  182}, \href{http://arxiv.org/abs/2104.05720}{{\tt arXiv:2104.05720
  [hep-ph]}}.

\bibitem{Bao:2022onq}
Y.~Bao, J.~Fan, and L.~Li, ``{Electroweak ALP searches at a muon collider},''
  \href{http://dx.doi.org/10.1007/JHEP08(2022)276}{{\em JHEP} {\bf 08} (2022)
  276}, \href{http://arxiv.org/abs/2203.04328}{{\tt arXiv:2203.04328
  [hep-ph]}}.

\bibitem{Han:2022mzp}
T.~Han, T.~Li, and X.~Wang, ``{Axion-Like Particles at High Energy Muon
  Colliders -- A White paper for Snowmass 2021},'' in {\em {Snowmass 2021}}.
\newblock 3, 2022.
\newblock \href{http://arxiv.org/abs/2203.05484}{{\tt arXiv:2203.05484
  [hep-ph]}}.

\bibitem{Li:2023tbx}
P.~Li, Z.~Liu, and K.-F. Lyu, ``{Heavy neutral leptons at muon colliders},''
  \href{http://dx.doi.org/10.1007/JHEP03(2023)231}{{\em JHEP} {\bf 03} (2023)
  231}, \href{http://arxiv.org/abs/2301.07117}{{\tt arXiv:2301.07117
  [hep-ph]}}.

\bibitem{Asadi:2023csb}
P.~Asadi, A.~Radick, and T.-T. Yu, ``{Interplay of freeze-in and freeze-out:
  Lepton-flavored dark matter and muon colliders},''
  \href{http://dx.doi.org/10.1103/PhysRevD.110.035022}{{\em Phys. Rev. D} {\bf
  110} (2024) no.~3, 035022}, \href{http://arxiv.org/abs/2312.03826}{{\tt
  arXiv:2312.03826 [hep-ph]}}.

\bibitem{Homiller:2022iax}
S.~Homiller, Q.~Lu, and M.~Reece, ``{Complementary signals of lepton flavor
  violation at a high-energy muon collider},''
  \href{http://dx.doi.org/10.1007/JHEP07(2022)036}{{\em JHEP} {\bf 07} (2022)
  036}, \href{http://arxiv.org/abs/2203.08825}{{\tt arXiv:2203.08825
  [hep-ph]}}.

\bibitem{Asadi:2024lfv}
P.~Asadi, H.~Bagherian, K.~Fraser, S.~Homiller, and Q.~Lu, ``{Lepton Flavor
  Violation: From Muon Decays to Muon Colliders},'' {\em to appear} (2024)  .

\bibitem{Cesarotti:2022ttv}
C.~Cesarotti, S.~Homiller, R.~K. Mishra, and M.~Reece, ``{Probing New Gauge
  Forces with a High-Energy Muon Beam Dump},''
  \href{http://dx.doi.org/10.1103/PhysRevLett.130.071803}{{\em Phys. Rev.
  Lett.} {\bf 130} (2023) no.~7, 071803},
  \href{http://arxiv.org/abs/2202.12302}{{\tt arXiv:2202.12302 [hep-ph]}}.

\bibitem{Cesarotti:2023sje}
C.~Cesarotti and R.~Gambhir, ``{The new physics case for beam-dump experiments
  with accelerated muon beams},''
  \href{http://dx.doi.org/10.1007/JHEP05(2024)283}{{\em JHEP} {\bf 05} (2024)
  283}, \href{http://arxiv.org/abs/2310.16110}{{\tt arXiv:2310.16110
  [hep-ph]}}.

\bibitem{Bogacz:2022xsj}
A.~Bogacz {\em et al.}, ``{The Physics Case for a Neutrino Factory},'' in {\em
  {Snowmass 2021}}.
\newblock March, 2022.
\newblock \href{http://arxiv.org/abs/2203.08094}{{\tt arXiv:2203.08094
  [hep-ph]}}.

\bibitem{Gori:2024zbs}
S.~Gori {\em et al.}, ``{ACE Science Workshop Report},''
  \href{http://arxiv.org/abs/2403.02422}{{\tt arXiv:2403.02422 [hep-ex]}}.

\bibitem{Altmannshofer:2014cla}
W.~Altmannshofer, P.~J. Fox, R.~Harnik, G.~D. Kribs, and N.~Raj, ``{Dark Matter
  Signals in Dilepton Production at Hadron Colliders},''
  \href{http://dx.doi.org/10.1103/PhysRevD.91.115006}{{\em Phys. Rev. D} {\bf
  91} (2015) no.~11, 115006}, \href{http://arxiv.org/abs/1411.6743}{{\tt
  arXiv:1411.6743 [hep-ph]}}.

\bibitem{Chway:2015lzg}
D.~Chway, R.~Derm\'\i{}\v{s}ek, T.~H. Jung, and H.~D. Kim, ``{Gluons to
  Diphotons via New Particles with Half the Signal\textquoteright{}s Invariant
  Mass},'' \href{http://dx.doi.org/10.1103/PhysRevLett.117.061801}{{\em Phys.
  Rev. Lett.} {\bf 117} (2016) no.~6, 061801},
  \href{http://arxiv.org/abs/1512.08221}{{\tt arXiv:1512.08221 [hep-ph]}}.

\bibitem{Mahbubani:2005pt}
R.~Mahbubani and L.~Senatore, ``{The Minimal model for dark matter and
  unification},'' \href{http://dx.doi.org/10.1103/PhysRevD.73.043510}{{\em
  Phys. Rev. D} {\bf 73} (2006)  043510},
  \href{http://arxiv.org/abs/hep-ph/0510064}{{\tt arXiv:hep-ph/0510064}}.

\bibitem{DEramo:2007anh}
F.~D'Eramo, ``{Dark matter and Higgs boson physics},''
  \href{http://dx.doi.org/10.1103/PhysRevD.76.083522}{{\em Phys. Rev. D} {\bf
  76} (2007)  083522}, \href{http://arxiv.org/abs/0705.4493}{{\tt
  arXiv:0705.4493 [hep-ph]}}.

\bibitem{Enberg:2007rp}
R.~Enberg, P.~J. Fox, L.~J. Hall, A.~Y. Papaioannou, and M.~Papucci, ``{LHC and
  dark matter signals of improved naturalness},''
  \href{http://dx.doi.org/10.1088/1126-6708/2007/11/014}{{\em JHEP} {\bf 11}
  (2007)  014}, \href{http://arxiv.org/abs/0706.0918}{{\tt arXiv:0706.0918
  [hep-ph]}}.

\bibitem{Cohen:2011ec}
T.~Cohen, J.~Kearney, A.~Pierce, and D.~Tucker-Smith, ``{Singlet-Doublet Dark
  Matter},'' \href{http://dx.doi.org/10.1103/PhysRevD.85.075003}{{\em Phys.
  Rev. D} {\bf 85} (2012)  075003}, \href{http://arxiv.org/abs/1109.2604}{{\tt
  arXiv:1109.2604 [hep-ph]}}.

\bibitem{Cheung:2013dua}
C.~Cheung and D.~Sanford, ``{Simplified Models of Mixed Dark Matter},''
  \href{http://dx.doi.org/10.1088/1475-7516/2014/02/011}{{\em JCAP} {\bf 02}
  (2014)  011}, \href{http://arxiv.org/abs/1311.5896}{{\tt arXiv:1311.5896
  [hep-ph]}}.

\bibitem{Abe:2014gua}
T.~Abe, R.~Kitano, and R.~Sato, ``{Discrimination of dark matter models in
  future experiments},''
  \href{http://dx.doi.org/10.1103/PhysRevD.91.095004}{{\em Phys. Rev. D} {\bf
  91} (2015) no.~9, 095004}, \href{http://arxiv.org/abs/1411.1335}{{\tt
  arXiv:1411.1335 [hep-ph]}}. [Erratum: Phys.Rev.D 96, 019902 (2017)].

\bibitem{Calibbi:2015nha}
L.~Calibbi, A.~Mariotti, and P.~Tziveloglou, ``{Singlet-Doublet Model: Dark
  matter searches and LHC constraints},''
  \href{http://dx.doi.org/10.1007/JHEP10(2015)116}{{\em JHEP} {\bf 10} (2015)
  116}, \href{http://arxiv.org/abs/1505.03867}{{\tt arXiv:1505.03867
  [hep-ph]}}.

\bibitem{Freitas:2015hsa}
A.~Freitas, S.~Westhoff, and J.~Zupan, ``{Integrating in the Higgs Portal to
  Fermion Dark Matter},'' \href{http://dx.doi.org/10.1007/JHEP09(2015)015}{{\em
  JHEP} {\bf 09} (2015)  015}, \href{http://arxiv.org/abs/1506.04149}{{\tt
  arXiv:1506.04149 [hep-ph]}}.

\bibitem{Banerjee:2016hsk}
S.~Banerjee, S.~Matsumoto, K.~Mukaida, and Y.-L.~S. Tsai, ``{WIMP Dark Matter
  in a Well-Tempered Regime: A case study on Singlet-Doublets Fermionic
  WIMP},'' \href{http://dx.doi.org/10.1007/JHEP11(2016)070}{{\em JHEP} {\bf 11}
  (2016)  070}, \href{http://arxiv.org/abs/1603.07387}{{\tt arXiv:1603.07387
  [hep-ph]}}.

\bibitem{Cai:2016sjz}
C.~Cai, Z.-H. Yu, and H.-H. Zhang, ``{CEPC Precision of Electroweak Oblique
  Parameters and Weakly Interacting Dark Matter: the Fermionic Case},''
  \href{http://dx.doi.org/10.1016/j.nuclphysb.2017.05.015}{{\em Nucl. Phys. B}
  {\bf 921} (2017)  181--210}, \href{http://arxiv.org/abs/1611.02186}{{\tt
  arXiv:1611.02186 [hep-ph]}}.

\bibitem{LopezHonorez:2017zrd}
L.~Lopez~Honorez, M.~H.~G. Tytgat, P.~Tziveloglou, and B.~Zaldivar, ``{On
  Minimal Dark Matter coupled to the Higgs},''
  \href{http://dx.doi.org/10.1007/JHEP04(2018)011}{{\em JHEP} {\bf 04} (2018)
  011}, \href{http://arxiv.org/abs/1711.08619}{{\tt arXiv:1711.08619
  [hep-ph]}}.

\bibitem{Fraser:2020dpy}
K.~Fraser, A.~Parikh, and W.~L. Xu, ``{A Closer Look at CP-Violating Higgs
  Portal Dark Matter as a Candidate for the GCE},''
  \href{http://dx.doi.org/10.1007/JHEP03(2021)123}{{\em JHEP} {\bf 03} (2021)
  123}, \href{http://arxiv.org/abs/2010.15129}{{\tt arXiv:2010.15129
  [hep-ph]}}.

\bibitem{Dreiner:2008tw}
H.~K. Dreiner, H.~E. Haber, and S.~P. Martin, ``{Two-component spinor
  techniques and Feynman rules for quantum field theory and supersymmetry},''
  \href{http://dx.doi.org/10.1016/j.physrep.2010.05.002}{{\em Phys. Rept.} {\bf
  494} (2010)  1--196}, \href{http://arxiv.org/abs/0812.1594}{{\tt
  arXiv:0812.1594 [hep-ph]}}.

\bibitem{Dedes:2014hga}
A.~Dedes and D.~Karamitros, ``{Doublet-Triplet Fermionic Dark Matter},''
  \href{http://dx.doi.org/10.1103/PhysRevD.89.115002}{{\em Phys. Rev. D} {\bf
  89} (2014) no.~11, 115002}, \href{http://arxiv.org/abs/1403.7744}{{\tt
  arXiv:1403.7744 [hep-ph]}}.

\bibitem{Giudice:1998xp}
G.~F. Giudice, M.~A. Luty, H.~Murayama, and R.~Rattazzi, ``{Gaugino mass
  without singlets},''
  \href{http://dx.doi.org/10.1088/1126-6708/1998/12/027}{{\em JHEP} {\bf 12}
  (1998)  027}, \href{http://arxiv.org/abs/hep-ph/9810442}{{\tt
  arXiv:hep-ph/9810442}}.

\bibitem{Wells:2003tf}
J.~D. Wells, ``{Implications of supersymmetry breaking with a little hierarchy
  between gauginos and scalars},'' in {\em {11th International Conference on
  Supersymmetry and the Unification of Fundamental Interactions}}.
\newblock 6, 2003.
\newblock \href{http://arxiv.org/abs/hep-ph/0306127}{{\tt
  arXiv:hep-ph/0306127}}.

\bibitem{Wells:2004di}
J.~D. Wells, ``{PeV-scale supersymmetry},''
  \href{http://dx.doi.org/10.1103/PhysRevD.71.015013}{{\em Phys. Rev. D} {\bf
  71} (2005)  015013}, \href{http://arxiv.org/abs/hep-ph/0411041}{{\tt
  arXiv:hep-ph/0411041}}.

\bibitem{Giudice:2011cg}
G.~F. Giudice and A.~Strumia, ``{Probing High-Scale and Split Supersymmetry
  with Higgs Mass Measurements},''
  \href{http://dx.doi.org/10.1016/j.nuclphysb.2012.01.001}{{\em Nucl. Phys. B}
  {\bf 858} (2012)  63--83}, \href{http://arxiv.org/abs/1108.6077}{{\tt
  arXiv:1108.6077 [hep-ph]}}.

\bibitem{Arvanitaki:2012ps}
A.~Arvanitaki, N.~Craig, S.~Dimopoulos, and G.~Villadoro, ``{Mini-Split},''
  \href{http://dx.doi.org/10.1007/JHEP02(2013)126}{{\em JHEP} {\bf 02} (2013)
  126}, \href{http://arxiv.org/abs/1210.0555}{{\tt arXiv:1210.0555 [hep-ph]}}.

\bibitem{Arkani-Hamed:2012fhg}
N.~Arkani-Hamed, A.~Gupta, D.~E. Kaplan, N.~Weiner, and T.~Zorawski, ``{Simply
  Unnatural Supersymmetry},'' \href{http://arxiv.org/abs/1212.6971}{{\tt
  arXiv:1212.6971 [hep-ph]}}.

\bibitem{Panico:2018hal}
G.~Panico, A.~Pomarol, and M.~Riembau, ``{EFT approach to the electron Electric
  Dipole Moment at the two-loop level},''
  \href{http://dx.doi.org/10.1007/JHEP04(2019)090}{{\em JHEP} {\bf 04} (2019)
  090}, \href{http://arxiv.org/abs/1810.09413}{{\tt arXiv:1810.09413
  [hep-ph]}}.

\bibitem{Atwood:1990cm}
D.~Atwood, C.~P. Burgess, C.~Hamazaou, B.~Irwin, and J.~A. Robinson, ``{One
  loop P and T odd W+- electromagnetic moments},''
  \href{http://dx.doi.org/10.1103/PhysRevD.42.3770}{{\em Phys. Rev. D} {\bf 42}
  (1990)  3770--3777}.

\bibitem{Kadoyoshi:1996bc}
T.~Kadoyoshi and N.~Oshimo, ``{Neutron electric dipole moment from
  supersymmetric anomalous W boson coupling},''
  \href{http://dx.doi.org/10.1103/PhysRevD.55.1481}{{\em Phys. Rev. D} {\bf 55}
  (1997)  1481--1486}, \href{http://arxiv.org/abs/hep-ph/9607301}{{\tt
  arXiv:hep-ph/9607301}}.

\bibitem{Chang:2005ac}
D.~Chang, W.-F. Chang, and W.-Y. Keung, ``{Electric dipole moment in the split
  supersymmetry models},''
  \href{http://dx.doi.org/10.1103/PhysRevD.71.076006}{{\em Phys. Rev. D} {\bf
  71} (2005)  076006}, \href{http://arxiv.org/abs/hep-ph/0503055}{{\tt
  arXiv:hep-ph/0503055}}.

\bibitem{Nakai:2016atk}
Y.~Nakai and M.~Reece, ``{Electric Dipole Moments in Natural Supersymmetry},''
  \href{http://dx.doi.org/10.1007/JHEP08(2017)031}{{\em JHEP} {\bf 08} (2017)
  031}, \href{http://arxiv.org/abs/1612.08090}{{\tt arXiv:1612.08090
  [hep-ph]}}.

\bibitem{Pilaftsis:2002fe}
A.~Pilaftsis, ``{Higgs mediated electric dipole moments in the MSSM: An
  application to baryogenesis and Higgs searches},''
  \href{http://dx.doi.org/10.1016/S0550-3213(02)00826-X}{{\em Nucl. Phys. B}
  {\bf 644} (2002)  263--289}, \href{http://arxiv.org/abs/hep-ph/0207277}{{\tt
  arXiv:hep-ph/0207277}}.

\bibitem{Giudice:2005rz}
G.~F. Giudice and A.~Romanino, ``{Electric dipole moments in split
  supersymmetry},''
  \href{http://dx.doi.org/10.1016/j.physletb.2006.01.027}{{\em Phys. Lett. B}
  {\bf 634} (2006)  307--314}, \href{http://arxiv.org/abs/hep-ph/0510197}{{\tt
  arXiv:hep-ph/0510197}}.

\bibitem{Arcadi:2017kky}
G.~Arcadi, M.~Dutra, P.~Ghosh, M.~Lindner, Y.~Mambrini, M.~Pierre, S.~Profumo,
  and F.~S. Queiroz, ``{The waning of the WIMP? A review of models, searches,
  and constraints},''
  \href{http://dx.doi.org/10.1140/epjc/s10052-018-5662-y}{{\em Eur. Phys. J. C}
  {\bf 78} (2018) no.~3, 203}, \href{http://arxiv.org/abs/1703.07364}{{\tt
  arXiv:1703.07364 [hep-ph]}}.

\bibitem{Roszkowski:2017nbc}
L.~Roszkowski, E.~M. Sessolo, and S.~Trojanowski, ``{WIMP dark matter
  candidates and searches\textemdash{}current status and future prospects},''
  \href{http://dx.doi.org/10.1088/1361-6633/aab913}{{\em Rept. Prog. Phys.}
  {\bf 81} (2018) no.~6, 066201}, \href{http://arxiv.org/abs/1707.06277}{{\tt
  arXiv:1707.06277 [hep-ph]}}.

\bibitem{Egana-Ugrinovic:2018roi}
D.~Egana-Ugrinovic, M.~Low, and J.~T. Ruderman, ``{Charged Fermions Below 100
  GeV},'' \href{http://dx.doi.org/10.1007/JHEP05(2018)012}{{\em JHEP} {\bf 05}
  (2018)  012}, \href{http://arxiv.org/abs/1801.05432}{{\tt arXiv:1801.05432
  [hep-ph]}}.

\bibitem{LEPSUSYWG1}
{\bf LEPSUSYWG, ALEPH, DELPHI, L3 and OPAL} Collaboration, ``{Combined lep
  chargino results, up to 208 gev for large m0},''
  \href{http://arxiv.org/abs/http://lepsusy.web.cern.ch/lepsusy/www/inos\_moriond01/charginos\_pub.html}{{\tt
  http://lepsusy.web.cern.ch/lepsusy/www/inos\_moriond01/charginos\_pub.html}}.

\bibitem{LEPSUSYWG2}
{\bf LEPSUSYWG, ALEPH, DELPHI, L3 and OPAL} Collaboration, ``{Combined lep
  chargino results, up to 208 gev for low dm},''
  \href{http://arxiv.org/abs/http://lepsusy.web.cern.ch/lepsusy/www/inoslowdmsummer02/charginolowdm\_pub.html}{{\tt
  http://lepsusy.web.cern.ch/lepsusy/www/inoslowdmsummer02/charginolowdm\_pub.html}}.

\bibitem{Ibe:2012sx}
M.~Ibe, S.~Matsumoto, and R.~Sato, ``{Mass Splitting between Charged and
  Neutral Winos at Two-Loop Level},''
  \href{http://dx.doi.org/10.1016/j.physletb.2013.03.015}{{\em Phys. Lett. B}
  {\bf 721} (2013)  252--260}, \href{http://arxiv.org/abs/1212.5989}{{\tt
  arXiv:1212.5989 [hep-ph]}}.

\bibitem{Canepa:2020ntc}
A.~Canepa, T.~Han, and X.~Wang, ``{The Search for Electroweakinos},''
  \href{http://dx.doi.org/10.1146/annurev-nucl-031020-121031}{{\em Ann. Rev.
  Nucl. Part. Sci.} {\bf 70} (2020)  425--454},
  \href{http://arxiv.org/abs/2003.05450}{{\tt arXiv:2003.05450 [hep-ph]}}.

\bibitem{CMS:2024gyw}
{\bf CMS} Collaboration, A.~Hayrapetyan {\em et al.}, ``{Combined search for
  electroweak production of winos, binos, higgsinos, and sleptons in
  proton-proton collisions at s=13\,\,TeV},''
  \href{http://dx.doi.org/10.1103/PhysRevD.109.112001}{{\em Phys. Rev. D} {\bf
  109} (2024) no.~11, 112001}, \href{http://arxiv.org/abs/2402.01888}{{\tt
  arXiv:2402.01888 [hep-ex]}}.

\bibitem{ATLAS:2021yqv}
{\bf ATLAS} Collaboration, G.~Aad {\em et al.}, ``{Search for charginos and
  neutralinos in final states with two boosted hadronically decaying bosons and
  missing transverse momentum in $pp$ collisions at $\sqrt {s}$ = 13 TeV with
  the ATLAS detector},''
  \href{http://dx.doi.org/10.1103/PhysRevD.104.112010}{{\em Phys. Rev. D} {\bf
  104} (2021) no.~11, 112010}, \href{http://arxiv.org/abs/2108.07586}{{\tt
  arXiv:2108.07586 [hep-ex]}}.

\bibitem{ATLAS:2022hbt}
{\bf ATLAS} Collaboration, G.~Aad {\em et al.}, ``{Search for direct pair
  production of sleptons and charginos decaying to two leptons and neutralinos
  with mass splittings near the W-boson mass in $ \sqrt{s} $ = 13 TeV pp
  collisions with the ATLAS detector},''
  \href{http://dx.doi.org/10.1007/JHEP06(2023)031}{{\em JHEP} {\bf 06} (2023)
  031}, \href{http://arxiv.org/abs/2209.13935}{{\tt arXiv:2209.13935
  [hep-ex]}}.

\bibitem{ATLAS:2022zwa}
{\bf ATLAS} Collaboration, G.~Aad {\em et al.}, ``{Searches for new phenomena
  in events with two leptons, jets, and missing transverse momentum in
  139~fb$^{-1}$ of ${\sqrt{s}}=13$ TeV ${pp}$ collisions with the ATLAS
  detector},'' \href{http://dx.doi.org/10.1140/epjc/s10052-023-11434-w}{{\em
  Eur. Phys. J. C} {\bf 83} (2023) no.~6, 515},
  \href{http://arxiv.org/abs/2204.13072}{{\tt arXiv:2204.13072 [hep-ex]}}.

\bibitem{Han:2020uak}
T.~Han, Z.~Liu, L.-T. Wang, and X.~Wang, ``{WIMPs at High Energy Muon
  Colliders},'' \href{http://dx.doi.org/10.1103/PhysRevD.103.075004}{{\em Phys.
  Rev. D} {\bf 103} (2021) no.~7, 075004},
  \href{http://arxiv.org/abs/2009.11287}{{\tt arXiv:2009.11287 [hep-ph]}}.

\bibitem{Capdevilla:2021fmj}
R.~Capdevilla, F.~Meloni, R.~Simoniello, and J.~Zurita, ``{Hunting wino and
  higgsino dark matter at the muon collider with disappearing tracks},''
  \href{http://dx.doi.org/10.1007/JHEP06(2021)133}{{\em JHEP} {\bf 06} (2021)
  133}, \href{http://arxiv.org/abs/2102.11292}{{\tt arXiv:2102.11292
  [hep-ph]}}.

\bibitem{Capdevilla:2024bwt}
R.~Capdevilla, F.~Meloni, and J.~Zurita, ``{Discovering Electroweak Interacting
  Dark Matter at Muon Colliders using Soft Tracks},''
  \href{http://arxiv.org/abs/2405.08858}{{\tt arXiv:2405.08858 [hep-ph]}}.

\bibitem{Peskin:1990zt}
M.~E. Peskin and T.~Takeuchi, ``{A New constraint on a strongly interacting
  Higgs sector},'' \href{http://dx.doi.org/10.1103/PhysRevLett.65.964}{{\em
  Phys. Rev. Lett.} {\bf 65} (1990)  964--967}.

\bibitem{Peskin:1991sw}
M.~E. Peskin and T.~Takeuchi, ``{Estimation of oblique electroweak
  corrections},'' \href{http://dx.doi.org/10.1103/PhysRevD.46.381}{{\em Phys.
  Rev. D} {\bf 46} (1992)  381--409}.

\bibitem{ParticleDataGroup:2024cfk}
{\bf Particle Data Group} Collaboration, S.~Navas {\em et al.}, ``{Review of
  particle physics},''
  \href{http://dx.doi.org/10.1103/PhysRevD.110.030001}{{\em Phys. Rev. D} {\bf
  110} (2024) no.~3, 030001}.

\bibitem{ATLAS:2022tnm}
{\bf ATLAS} Collaboration, G.~Aad {\em et al.}, ``{Measurement of the
  properties of Higgs boson production at $\sqrt{s} = 13$ TeV in the
  $H\to\gamma\gamma$ channel using $139$ fb$^{-1}$ of $pp$ collision data with
  the ATLAS experiment},''
  \href{http://dx.doi.org/10.1007/JHEP07(2023)088}{{\em JHEP} {\bf 07} (2023)
  088}, \href{http://arxiv.org/abs/2207.00348}{{\tt arXiv:2207.00348
  [hep-ex]}}.

\bibitem{CMS:2021kom}
{\bf CMS} Collaboration, A.~M. Sirunyan {\em et al.}, ``{Measurements of Higgs
  boson production cross sections and couplings in the diphoton decay channel
  at $ \sqrt{\mathrm{s}} $ = 13 TeV},''
  \href{http://dx.doi.org/10.1007/JHEP07(2021)027}{{\em JHEP} {\bf 07} (2021)
  027}, \href{http://arxiv.org/abs/2103.06956}{{\tt arXiv:2103.06956
  [hep-ex]}}.

\bibitem{ATLAS:2023yqk}
{\bf ATLAS, CMS} Collaboration, G.~Aad {\em et al.}, ``{Evidence for the Higgs
  Boson Decay to a Z Boson and a Photon at the LHC},''
  \href{http://dx.doi.org/10.1103/PhysRevLett.132.021803}{{\em Phys. Rev.
  Lett.} {\bf 132} (2024) no.~2, 021803},
  \href{http://arxiv.org/abs/2309.03501}{{\tt arXiv:2309.03501 [hep-ex]}}.

\bibitem{Cepeda:2019klc}
M.~Cepeda {\em et al.}, ``{Report from Working Group 2}: {Higgs Physics at the
  HL-LHC and HE-LHC},''
  \href{http://dx.doi.org/10.23731/CYRM-2019-007.221}{{\em CERN Yellow Rep.
  Monogr.} {\bf 7} (2019)  221--584},
  \href{http://arxiv.org/abs/1902.00134}{{\tt arXiv:1902.00134 [hep-ph]}}.

\bibitem{ATLAS:2016neq}
{\bf ATLAS, CMS} Collaboration, G.~Aad {\em et al.}, ``{Measurements of the
  Higgs boson production and decay rates and constraints on its couplings from
  a combined ATLAS and CMS analysis of the LHC pp collision data at $
  \sqrt{s}=7 $ and 8 TeV},''
  \href{http://dx.doi.org/10.1007/JHEP08(2016)045}{{\em JHEP} {\bf 08} (2016)
  045}, \href{http://arxiv.org/abs/1606.02266}{{\tt arXiv:1606.02266
  [hep-ex]}}.

\bibitem{ATLAS:2020rej}
{\bf ATLAS} Collaboration, G.~Aad {\em et al.}, ``{Higgs boson production
  cross-section measurements and their EFT interpretation in the $4\ell $ decay
  channel at $\sqrt{s}=$13 TeV with the ATLAS detector},''
  \href{http://dx.doi.org/10.1140/epjc/s10052-020-8227-9}{{\em Eur. Phys. J. C}
  {\bf 80} (2020) no.~10, 957}, \href{http://arxiv.org/abs/2004.03447}{{\tt
  arXiv:2004.03447 [hep-ex]}}. [Erratum: Eur.Phys.J.C 81, 29 (2021), Erratum:
  Eur.Phys.J.C 81, 398 (2021)].

\bibitem{CMS:2022dwd}
{\bf CMS} Collaboration, A.~Tumasyan {\em et al.}, ``{A portrait of the Higgs
  boson by the CMS experiment ten years after the discovery.},''
  \href{http://dx.doi.org/10.1038/s41586-022-04892-x}{{\em Nature} {\bf 607}
  (2022) no.~7917, 60--68}, \href{http://arxiv.org/abs/2207.00043}{{\tt
  arXiv:2207.00043 [hep-ex]}}. [Erratum: Nature 623, (2023)].

\bibitem{Chiesa:2021qpr}
M.~Chiesa, B.~Mele, and F.~Piccinini, ``{Multi Higgs production via photon
  fusion at future multi-TeV muon colliders},''
  \href{http://dx.doi.org/10.1140/epjc/s10052-024-12882-8}{{\em Eur. Phys. J.
  C} {\bf 84} (2024) no.~5, 543}, \href{http://arxiv.org/abs/2109.10109}{{\tt
  arXiv:2109.10109 [hep-ph]}}.

\bibitem{Hahn:1998yk}
T.~Hahn and M.~Perez-Victoria, ``{Automatized one loop calculations in
  four-dimensions and D-dimensions},''
  \href{http://dx.doi.org/10.1016/S0010-4655(98)00173-8}{{\em Comput. Phys.
  Commun.} {\bf 118} (1999)  153--165},
  \href{http://arxiv.org/abs/hep-ph/9807565}{{\tt arXiv:hep-ph/9807565}}.

\bibitem{Hahn:2000kx}
T.~Hahn, ``{Generating Feynman diagrams and amplitudes with FeynArts 3},''
  \href{http://dx.doi.org/10.1016/S0010-4655(01)00290-9}{{\em Comput. Phys.
  Commun.} {\bf 140} (2001)  418--431},
  \href{http://arxiv.org/abs/hep-ph/0012260}{{\tt arXiv:hep-ph/0012260}}.

\bibitem{Aoki:1980hh}
K.-i. Aoki, Z.~Hioki, R.~Kawabe, M.~Konuma, and T.~Muta, ``{One Loop
  Corrections to Neutrino $e$ Scattering in {Weinberg-Salam} Theory: Neutral
  Current Processes},'' \href{http://dx.doi.org/10.1143/PTP.64.707}{{\em Prog.
  Theor. Phys.} {\bf 64} (1980)  707}.

\bibitem{Aoki:1980ix}
K.-i. Aoki, Z.~Hioki, R.~Kawabe, M.~Konuma, and T.~Muta, ``{Electroweak
  Radiative Corrections to High-Energy $\nu e$ Scatterings},''
  \href{http://dx.doi.org/10.1143/PTP.65.1001}{{\em Prog. Theor. Phys.} {\bf
  65} (1981)  1001}.

\bibitem{Aoki:1982ed}
K.~I. Aoki, Z.~Hioki, M.~Konuma, R.~Kawabe, and T.~Muta, ``{Electroweak Theory.
  Framework of On-Shell Renormalization and Study of Higher Order Effects},''
  \href{http://dx.doi.org/10.1143/PTPS.73.1}{{\em Prog. Theor. Phys. Suppl.}
  {\bf 73} (1982)  1--225}.

\bibitem{Denner:1991kt}
A.~Denner, ``{Techniques for calculation of electroweak radiative corrections
  at the one loop level and results for W physics at LEP-200},''
  \href{http://dx.doi.org/10.1002/prop.2190410402}{{\em Fortsch. Phys.} {\bf
  41} (1993)  307--420}, \href{http://arxiv.org/abs/0709.1075}{{\tt
  arXiv:0709.1075 [hep-ph]}}.

\bibitem{Costantini:2020stv}
A.~Costantini, F.~De~Lillo, F.~Maltoni, L.~Mantani, O.~Mattelaer, R.~Ruiz, and
  X.~Zhao, ``{Vector boson fusion at multi-TeV muon colliders},''
  \href{http://dx.doi.org/10.1007/JHEP09(2020)080}{{\em JHEP} {\bf 09} (2020)
  080}, \href{http://arxiv.org/abs/2005.10289}{{\tt arXiv:2005.10289
  [hep-ph]}}.

\bibitem{Ruiz:2021tdt}
R.~Ruiz, A.~Costantini, F.~Maltoni, and O.~Mattelaer, ``{The Effective Vector
  Boson Approximation in high-energy muon collisions},''
  \href{http://dx.doi.org/10.1007/JHEP06(2022)114}{{\em JHEP} {\bf 06} (2022)
  114}, \href{http://arxiv.org/abs/2111.02442}{{\tt arXiv:2111.02442
  [hep-ph]}}.

\bibitem{Garosi:2023bvq}
F.~Garosi, D.~Marzocca, and S.~Trifinopoulos, ``{LePDF: Standard Model PDFs for
  high-energy lepton colliders},''
  \href{http://dx.doi.org/10.1007/JHEP09(2023)107}{{\em JHEP} {\bf 09} (2023)
  107}, \href{http://arxiv.org/abs/2303.16964}{{\tt arXiv:2303.16964
  [hep-ph]}}.

\bibitem{Cutkosky:1960sp}
R.~E. Cutkosky, ``{Singularities and discontinuities of Feynman amplitudes},''
  \href{http://dx.doi.org/10.1063/1.1703676}{{\em J. Math. Phys.} {\bf 1}
  (1960)  429--433}.

\bibitem{Remiddi:1981hn}
E.~Remiddi, ``{Dispersion Relations for Feynman Graphs},'' {\em Helv. Phys.
  Acta} {\bf 54} (1982)  364.

\bibitem{Aloni:2021wzk}
D.~Aloni, P.~Asadi, Y.~Nakai, M.~Reece, and M.~Suzuki, ``{Spontaneous CP
  violation and horizontal symmetry in the MSSM: toward lepton flavor
  naturalness},'' \href{http://dx.doi.org/10.1007/JHEP09(2021)031}{{\em JHEP}
  {\bf 09} (2021)  031}, \href{http://arxiv.org/abs/2104.02679}{{\tt
  arXiv:2104.02679 [hep-ph]}}.

\bibitem{Wilks:1938dza}
S.~S. Wilks, ``{The Large-Sample Distribution of the Likelihood Ratio for
  Testing Composite Hypotheses},''
  \href{http://dx.doi.org/10.1214/aoms/1177732360}{{\em Annals Math. Statist.}
  {\bf 9} (1938) no.~1, 60--62}.

\bibitem{Cowan:2010js}
G.~Cowan, K.~Cranmer, E.~Gross, and O.~Vitells, ``{Asymptotic formulae for
  likelihood-based tests of new physics},''
  \href{http://dx.doi.org/10.1140/epjc/s10052-011-1554-0}{{\em Eur. Phys. J. C}
  {\bf 71} (2011)  1554}, \href{http://arxiv.org/abs/1007.1727}{{\tt
  arXiv:1007.1727 [physics.data-an]}}. [Erratum: Eur.Phys.J.C 73, 2501 (2013)].

\bibitem{Hamada:2022mua}
Y.~Hamada, R.~Kitano, R.~Matsudo, H.~Takaura, and M.~Yoshida,
  ``{$\mu$TRISTAN},'' \href{http://dx.doi.org/10.1093/ptep/ptac059}{{\em PTEP}
  {\bf 2022} (2022) no.~5, 053B02}, \href{http://arxiv.org/abs/2201.06664}{{\tt
  arXiv:2201.06664 [hep-ph]}}.

\end{thebibliography}\endgroup
\bibliographystyle{utphys}
}

\end{document}